\documentclass[twocolumn]{aastex63} 

\usepackage{graphicx} 
\usepackage{float} 
\usepackage{wrapfig} 
\usepackage{lipsum} 
\usepackage{hyperref}
\usepackage{times}
\usepackage{amsmath}
\usepackage{graphicx}
\usepackage{subfigure}
\usepackage{placeins}
\usepackage{hyperref}
\usepackage{gensymb}
\usepackage{upgreek}
\usepackage{natbib}

\usepackage{graphicx}
\usepackage{subfigure}
\usepackage{multirow}
\usepackage{comment}
\usepackage{natbib}
\usepackage{hyperref}
\usepackage{mathtools}
\usepackage{mathrsfs}
\usepackage{fontenc}
\usepackage{color}
\usepackage{url}
\usepackage{hyperref}
\usepackage{gensymb}
\usepackage{pifont}
\graphicspath{{./}}

\newcommand       \K            {\,{\rm K}}
\newcommand       \mum          {{\rm \mu m}}

\usepackage{afterpage}

\shorttitle{Feedback Effects in Low-luminosity AGN}
\shortauthors{Zhang et al.}

\begin{document}

\title{Evidence of Feedback Effects in Low-luminosity Active Galactic Nuclei Revealed by JWST Spectroscopy}

\author[0000-0003-4937-9077]{Lulu Zhang}
\affiliation{The University of Texas at San Antonio, One UTSA Circle, San Antonio, TX 78249, USA; lulu.zhang@utsa.edu; l.l.zhangastro@gmail.com}

\author[0000-0001-7827-5758]{Chris Packham}
\affiliation{The University of Texas at San Antonio, One UTSA Circle, San Antonio, TX 78249, USA; lulu.zhang@utsa.edu; l.l.zhangastro@gmail.com}
\affiliation{National Astronomical Observatory of Japan, National Institutes of Natural Sciences (NINS), 2-21-1 Osawa, Mitaka, Tokyo 181-8588, Japan}

\author[0000-0002-4457-5733]{Erin K. S. Hicks}
\affiliation{Department of Physics and Astronomy, University of Alaska Anchorage, Anchorage, AK 99508-4664, USA}
\affiliation{The University of Texas at San Antonio, One UTSA Circle, San Antonio, TX 78249, USA; lulu.zhang@utsa.edu; l.l.zhangastro@gmail.com}
\affiliation{Department of Physics, University of Alaska, Fairbanks, Alaska 99775-5920, USA}

\author[0000-0003-4949-7217]{Ric I. Davies}
\affiliation{Max-Planck-Institut für extraterrestrische Physik, Postfach 1312, D-85741, Garching, Germany}

\author[0009-0007-6992-2555]{Daniel E. Delaney}
\affiliation{Department of Physics, University of Alaska, Fairbanks, Alaska 99775-5920, USA}
\affiliation{Department of Physics and Astronomy, University of Alaska Anchorage, Anchorage, AK 99508-4664, USA}

\author[0000-0003-2658-7893]{Francoise Combes}
\affiliation{LUX, Observatoire de Paris, Coll{\`e}ge de France, PSL University, CNRS, Sorbonne University, Paris}

\author[0000-0002-4005-9619]{Miguel Pereira-Santaella}
\affiliation{Instituto de F{\'i}sica Fundamental, CSIC, Calle Serrano 123, 28006 Madrid, Spain}

\author[0000-0001-6794-2519]{Almudena Alonso-Herrero}
\affiliation{Centro de Astrobiolog\'{\i}a (CAB), CSIC-INTA, Camino Bajo del Castillo s/n, E-28692 Villanueva de la Ca\~nada, Madrid, Spain}

\author[0000-0001-5231-2645]{Claudio Ricci}
\affiliation{Department of Astronomy, University of Geneva, ch. d'Ecogia 16, 1290, Versoix, Switzerland}
\affiliation{ Instituto de Estudios Astrof\'isicos, Facultad de Ingenier\'ia y Ciencias, Universidad Diego Portales, Av. Ej\'ercito Libertador 441, Santiago, Chile}

\author[0000-0002-2356-8358]{Omaira Gonz{\'a}lez-Mart{\'i}n}
\affiliation{Instituto de Radioastronom{\'i}a and Astrof{\'i}sica (IRyA-UNAM), 3-72 (Xangari), 8701, Morelia, Mexico}

\author[0000-0002-9610-0123]{Laura Hermosa Mu{\~n}oz}
\affiliation{Centro de Astrobiolog\'{\i}a (CAB), CSIC-INTA, Camino Bajo del Castillo s/n, E-28692 Villanueva de la Ca\~nada, Madrid, Spain}

\author[0000-0002-9627-5281]{Ismael Garc{\'i}a-Bernete}
\affiliation{Centro de Astrobiolog\'{\i}a (CAB), CSIC-INTA, Camino Bajo del Castillo s/n, E-28692 Villanueva de la Ca\~nada, Madrid, Spain}

\author[0000-0001-8353-649X]{Cristina Ramos Almeida}
\affiliation{Instituto de Astrof{\'i}sica de Canarias, Calle V{\'i}a L{\'a}ctea, s/n, E-38205, La Laguna, Tenerife, Spain}
\affiliation{Departamento de Astrof{\'i}sica, Universidad de La Laguna, E-38206, La Laguna, Tenerife, Spain}

\author[0000-0001-6854-7545]{Dimitra Rigopoulou}
\affiliation{Department of Physics, University of Oxford, Keble Road, Oxford OX1 3RH, UK}
\affiliation{School of Sciences, European University Cyprus, Diogenes street, Engomi, 1516 Nicosia, Cyprus}

\author[0000-0002-6460-3682]{Fergus R. Donnan}
\affiliation{Department of Astronomy and Astrophysics, University of California, San Diego, La Jolla, CA 92093, USA}

\author[0000-0001-9791-4228]{Enrica Bellocchi}
\affiliation{Departamento de F\'isica de la Tierra y Astrof\'isica, Fac. de CC. F\'isicas, Universidad Complutense de Madrid, 28040 Madrid, Spain}
\affiliation{Instituto de F\'isica de Part\'iculas y del Cosmos IPARCOS, Fac. CC. F\'isicas, Universidad Complutense de Madrid, 28040 Madrid, Spain}

\author[0000-0003-4209-639X]{Nancy A. Levenson}
\affiliation{Space Telescope Science Institute, 3700 San Martin Drive Baltimore, Maryland 21218, USA}

\author[0000-0003-1810-0889]{Martin J. Ward}
\affiliation{Centre for Extragalactic Astronomy, Durham University, South Road, Durham DH1 3LE, UK}

\author[0000-0003-0444-6897]{Santiago Garc{\'i}a-Burillo}
\affiliation{Observatorio Astron{\'o}mico Nacional (OAN-IGN)-Observatorio de Madrid, Alfonso XII, 3, 28014, Madrid, Spain}

\author[0000-0002-6353-1111]{Sebastian F. Hoenig}
\affiliation{School of Physics and Astronomy, University of Southampton, Southampton SO17 1BJ, UK}



\begin{abstract}

This letter presents an analysis of the infrared ($\sim 3-28\,\mum$) spectra extracted from the nuclear ($r < 150$ pc) regions of four low-luminosity active galactic nuclei (AGN), observed by JWST NIRSpec/IFU and MIRI/MRS as an extension of the Galaxy Activity, Torus, and Outflow Survey (GATOS). We find that, compared to higher-luminosity AGN, these low-luminosity AGN exhibit distinct properties in their emission of ionized gas, polycyclic aromatic hydrocarbons (PAHs), and molecular hydrogen (H$_2$). Specifically, the low-luminosity AGN exhibit relatively weak high-ionization potential lines (e.g., [Ne~{\small V}] and [O~{\small IV}]), and the line ratios suggest that fast radiative shocks (with $v_{\rm s}$ of $\sim \rm 100s\,km\,s^{-1}$) are the primary excitation source of ionized gas therein. Under the low-excitation conditions of their nuclear regions, these low-luminosity AGN generally exhibit a higher fraction of PAHs with large size ($N_{\rm C} \gtrsim 200$), reflecting the preferential destruction of smaller PAH molecules by AGN feedback. Furthermore, the H$_2$ transitions in these low-luminosity AGN are not fully thermalized, with slow, plausibly jet-driven molecular shocks (with $v_{\rm s} \leq \rm 10\,km\,s^{-1}$) likely being the extra excitation source. Taken together with results from the literature, these findings indicate that feedback operates in both low- and high-luminosity AGN, albeit its impact varies with AGN luminosity. In particular, systematic variations in PAH band ratios are found across AGN, demonstrating the differing influence of feedback in AGN of varying luminosities and highlighting the potential of PAH band ratios as diagnostics for distinguishing kinetic- and radiative-mode AGN feedback.

\end{abstract}

\keywords{galaxies: active galactic nucleus --- galaxies: ISM --- infrared: ISM --- galaxies: star formation}

\section{Introduction}

Widely regarded to coevolve with their host galaxies (e.g., \citealt{Magorrian.etal.1998, Ferrarese&Merritt2000, Gebhardt.etal.2000}; and see review \citealt{Kormendy&Ho2013}), supermassive black holes (SMBHs) reside in most, if not all, massive galaxies and can influence their hosts and large-scale surroundings in a variety of ways (e.g., \citealt{Schaye.etal.2015, Pillepich.etal.2018, Dave.etal.2019}; and see reviews \citealt{Fabian2012, Heckman&Best2014, Harrison&RamosAlmeida2024}). In the local universe, the most prevalent accreting SMBHs are low-luminosity active galactic nuclei (AGN), often manifesting as low-ionization nuclear emission-line regions (LINERs; \citealt{Heckman1980}) characterized by very low radiative efficiency, faint luminosities, and more pronounced jet-like features (e.g., \citealt{Ho.etal.1993, Ho.etal.2003, Ho1999, Ho2009, DiMatteo.etal.2003, Pellegrini.etal.2003, Mason.etal.2012, Fernandez-Ontiveros.etal.2023}; and see review \citealt{Ho2008}). In contrast to vigorously accreting, high-luminosity AGN that can expel gas from their host galaxies through {\it the radiative (or quasar) mode} feedback (e.g., winds/outflows; \citealt{DiMatteo.etal.2005, Hopkins.etal.2008a, Harrison.etal.2014}), these highly sub-Eddington, low-luminosity AGN can in theory redirect most of their accretion power from radiation to kinetic energy (see reviews \citealt{McNamara&Nulsen2007, Yuan&Narayan2014}), and interact with their surroundings mainly via {\it the kinetic (or radio/jet) mode} feedback (e.g., jets/shocks; \citealt{Weinberger.etal.2017, Dave.etal.2019}). 

The kinetic mode feedback has attracted considerable attention given its crucial role in sustaining the quenching phase of massive bulges and elliptical galaxies (see reviews  \citealt{Fabian2012, Heckman&Best2014}). These systems are theoretically regarded as the ultimate fate of galaxy evolution and are of great importance for our understanding of AGN--host co-evolution (\citealt{Hopkins.etal.2006, Hopkins.etal.2008a, Hopkins.etal.2008b}). Given their significance in the evolutionary pathway of galaxies, the kinetic mode feedback has been widely studied through numerical simulations and optical/X-ray/radio observations (e.g., \citealt{DiMatteo.etal.2003, Nesvadba.etal.2017, Pillepich.etal.2018, Dave.etal.2020, Fernandez-Ontiveros.etal.2023, Wang.etal.2024, Ilha.etal.2025}). JWST (\citealt{Gardner.etal.2023}), with its superb sensitivity and broad wavelength coverage, offers new opportunities to probe the specific effects of kinetic-mode AGN feedback through the rich information on gas content, kinematics, and excitation provided by abundant infrared diagnostics relatively immune from dust extinction, including a broad suite of ionized emission lines, H$_2$ transitions, and PAH features (e.g., \citealt{Pereira-Santaella.etal.2022, Garcia-Bernete.etal.2022a, Garcia-Bernete.etal.2024, Donnan.etal.2023, Hernandez.etal.2023, Zhang&Ho2023, Chown.etal.2024, Goold.etal.2024,  HermosaMunoz.etal.2025a, HermosaMunoz.etal.2025b, Peeters.etal.2024, Zhang.etal.2024a, Zhang.etal.2024b, AlonsoHerrero.etal.2025, Ceci.etal.2025, Lopez.etal.2025, Ogle.etal.2025, RamosAlmeida.etal.2025}).

Combined with theoretical calculations (e.g., \citealt{Kristensen.etal.2023, Rigopoulou.etal.2024, Lopez.etal.2025, Zhang.etal.2025}), infrared emission features of ionized gas, H$_2$, and PAHs together can provide strong constraints on the physical conditions around AGN and hence underlying feedback mechanisms. In particular, the intrinsic and relative intensities of individual PAH features at different wavelengths (e.g., 3.3, 6.2, 7.7, 8.6, 11.3, 12.7, and 17.0 $\mum$) vary markedly across different galactic environments (e.g., \citealt{Genzel.etal.1998, Peeters.etal.2002, Kaneda.etal.2005, Farrah.etal.2007, Gordon.etal.2008, Hunt.etal.2010, Lebouteiller.etal.2011}), and the variation is particularly pronounced in the hostile environments around AGN (e.g., \citealt{Smith.etal.2007, ODowd.etal.2009, Diamond-Stanic&Rieke2010, Sales.etal.2010, Garcia-Bernete.etal.2022a, Garcia-Bernete.etal.2022b, Garcia-Bernete.etal.2024, Zhang.etal.2022, Zhang.etal.2024b, Lai.etal.2023, Zhang&Ho2023, Donnelly.etal.2024}). Specifically, low-luminosity AGN, especially LINERs, exhibit on kilo-parsec (kpc) scales relatively lower PAH 6.2 $\mu$m/7.7 $\mu$m and higher 11.3 $\mu$m/7.7 $\mu$m ratios compared to star-forming galaxies (SFGs), which can be attributed to the preferential destruction of the smaller PAHs by shocks within the overall PAH population (\citealt{Zhang.etal.2022}). In addition, low-redshift quasars, where the radiative mode feedback is likely to dominate, exhibit even lower PAH 6.2 $\mu$m/7.7 $\mu$m but also lower 11.3 $\mu$m/7.7 $\mu$m ratios (\citealt{Xie&Ho2022}). The distinct PAH characteristics observed in low- and high-luminosity AGN provide preliminary evidence that PAH features may serve as useful diagnostics for distinguishing between different feedback modes.

This letter is the first in a series designed to investigate the kinetic mode feedback in a sample of four low-luminosity AGN, through analysis of infrared emission features obtained from dedicated and archival JWST spectroscopic observations. Low-luminosity AGN are targeted here because kinetic-mode feedback is expected to be prevalent in these systems as outlined above, although growing evidence indicates that such feedback can also be important in high-luminosity AGN (e.g., \citealt{Ilha.etal.2025, Roy.etal.2025, Vayner.etal.2025}). This first paper focuses on the analysis of the infrared ($\sim 3-28\,\mum$) spectra extracted from the nuclear ($r < 150$ pc) regions of the sample, while the spatially resolved analysis over a larger field of view (FoV) will be presented in subsequent papers. After introducing the targets and observations, as well as the data reduction, spectral extraction, and decomposition in Section~\ref{sec2}, we present the main results on the distinct emission properties of ionized gas, PAHs, and H$_2$ in the nuclear regions of the four low-luminosity AGN in Section~\ref{sec3}. We further discuss the underlying mechanisms responsible for the systematically distributed PAH band ratios observed across diverse galactic systems in Section~\ref{sec4}. A summary of the main points of this letter is presented in Section~\ref{sec5}.

\section{Observation and Analysis}\label{sec2}

\subsection{Targets and Observations}\label{sec2.1}

The sample analyzed here consists of four low-luminosity AGN selected by JWST Cycle 3 General Observer (GO) Program 4972 (PI: L. Zhang) from the Spitzer Infrared Nearby Galaxies Survey (SINGS; \citealt{Kennicutt.etal.2003}). The SINGS survey is a comprehensive infrared imaging and (mapping-mode) spectroscopic survey by the Spitzer with a rich repository of ancillary multi-band data sets (see e.g., \citealt{Dale.etal.2006, Smith.etal.2007,Moustakas.etal.2010}). GO Program 4972 is among the first JWST programs dedicated to low-luminosity AGN, and it is the first designed explicitly to assess the effectiveness of PAH features as diagnostics of the kinetic mode feedback. To this end, we proposed to investigate the spatially resolved characteristics of PAH features and other infrared emission lines in the $\sim 3 - 28\,\mum$ wavelength range around these low-luminosity AGN. Specifically, this sample includes all SINGS AGN that are classified as LINERs in the optical band (\citealt{Smith.etal.2007}) and also exhibit prominent PAH features in their Spitzer/IRS spectra extracted from the central $10\arcsec$ regions (\citealt{Zhang.etal.2022}). Additionally, all four LINERs show evidence of mechanical processes (radio cores and/or shocked gas) associated with their nuclei (e.g., \citealt{Nagar.etal.2002, Nagar.etal.2005, Nemmen.etal.2006, Nemmen.etal.2014, Nyland.etal.2013, Pellegrini.etal.2013, Mezcua&Prieto2014}). Note that JWST spectroscopy of NGC~4736 were already included in JWST Cycle 1 GO Program 2016 (PI: A. Seth; \citealt{Goold.etal.2024}), GO Program 4972 therefore only requested JWST spectroscopy for NGC~1097, NGC~1266, and NGC~3190. See Table~\ref{tabinfo} for basic properties of the sample.

\startlongtable
\setlength{\tabcolsep}{6pt}
\begin{deluxetable*}{ccccccccc}
\tablecolumns{9}
\tablecaption{Basic Properties of the Sampled Low-luminosity AGN}
\tablehead{
\colhead{Target} & \colhead{Hubble Type} & \colhead{\,\,\,\,\,\,\,$z$\,\,\,\,\,\,\,} & \colhead{\,\,\,\,\,\,\,$D$\,\,\,\,\,\,\,} & \colhead{\,\,\,\,\,\,\,$i$\,\,\,\,\,\,\,} & \colhead{log $L^{\rm Xray}_{\rm (2-10)keV}$} & \colhead{log $L^{\rm radio}_{\rm 1.4GHz}$} & \colhead{log $M_{\rm BH}$} & \colhead{log $\lambda_{\rm Edd}$} \\
\colhead{(-)} & \colhead{(-)} & \colhead{(-)} & \colhead{(Mpc)} & \colhead{($\degree$)} & \colhead{[$\rm erg\,s^{-1}$]} & \colhead{[$\rm W\,Hz^{-1}$]} & \colhead{[$\rm M_{\odot}$]} & \colhead{(-)} \\
\colhead{(1)} & \colhead{(2)} & \colhead{(3)} & \colhead{(4)} & \colhead{(5)} & \colhead{(6)} & \colhead{(7)} & \colhead{(8)} & \colhead{(9)} }
\startdata
NGC~1097$^{\dag}$ & SBb & 0.004240 & 17.2 & 35 & 40.8 & 22.2 & 8.1 & $-$4.2\\
NGC~1266$^{\ddag}$ & SB0& 0.007238 & 29.4 & 50 & 40.6 & 22.0 & 6.2 & $-$2.5\\
NGC~3190 & Sa& 0.004370 & 24.5 & 83 & 39.6 & 21.5 & 8.2 & $-$5.5\\
NGC~4736 & Sab & 0.001027 & 5.1 & 35 & 38.6 & 20.9 & 7.2 & $-$5.5 \\
\enddata
\tablecomments{\footnotesize Column (1): Target names; Column (2-5): Hubble types, redshifts, redshift-independent distances, and disk inclinations taken from the NASA/IPAC Extragalactic Database (NED); Column (6): Nuclear 2--10 keV X-ray luminosities taken from \cite{Ho2009} unless otherwise specified (the values in this table have been adjusted according to the adopted distances); Column (7): Nuclear 1.4 GHz radio luminosity densities derived from VLA flux densities retrieved from the NED. Column (8): Black hole masses derived from the $M_{\rm BH}-\sigma_{*}$ relation of \cite{Greene.etal.2020} unless otherwise specified, with $\sigma_{*}$ from \cite{Ho.etal.2009}. Column (9): Eddington ratios $\lambda_{\rm Edd} = L_{\rm bol}/L_{\rm Edd}$, with $L_{\rm bol} = 15.8\times L^{\rm Xray}_{\rm (2-10)keV}$ and $L_{\rm Edd} = 1.26\times10^{38}\ (M_{\rm BH}/{\rm M_{\odot}})$. [\dag] $L^{\rm Xray}_{\rm (2-10)keV}$ and $M_{\rm BH}$ are taken from \cite{Cisternas.etal.2013}. [\ddag] $L^{\rm Xray}_{\rm (2-10)keV}$ and $M_{\rm BH}$ are taken from \cite{Chen.etal.2023} and \cite{Alatalo.etal.2015}, respectively.}
\label{tabinfo}
\end{deluxetable*}

NGC~1097, NGC~1266, and NGC~3190 were observed under GO Program 4972 between November 2024 and May 2025 with JWST NIRSpec/IFU (\citealt{Boker.etal.2022, Jakobsen.etal.2022}) using the G395H/F290LP grating/filter combination and with MIRI/MRS (\citealt{Wells.etal.2015, Wright.etal.2023}) covering all four channels. JWST spectroscopic observations of NGC~4736 (including the above instrumental combinations) were carried out in June 2024. The two programs both adopted a 4-point dither patten for NIRSpec/IFU and MIRI/MRS exposures of the targets, focusing on their most central $\sim 300-600$ pc regions, where AGN activity dominates (\citealt{Moustakas.etal.2010}). The total on source exposure times range from 584 -- 875s and 777 -- 999s for NIRSpec/IFU and MIRI/MRS observations using NRSIRS2RAPID and FASTR1 readout patterns, respectively. In addition, GO Program 4972 adopted a recommended 2-point dither patten while GO Program 2016 used a 1-point non-dither patten for the corresponding MIRI/MRS background observations. Both programs also adopted a 1-point non-dither pattern for the NIRSpec/IFU leakage corrections. More details about the observational configurations are included in the program information webpages of programs \href{https://www.stsci.edu/jwst/science-execution/program-information?id=4972}{GO 4972} and \href{https://www.stsci.edu/jwst/science-execution/program-information?id=2016}{GO 2016}. The observations of the four targets can be accessed via doi: \href{https://archive.stsci.edu/doi/resolve/resolve.html?doi=10.17909/psw5-s410}{10.17909/psw5-s410}.

\subsection{Data Processing and Nuclear Spectrum Extraction}\label{sec2.2}

The data were reduced using the JWST Science Calibration Pipeline (v1.18.0; \citealt{Bushouse.etal.2025}) with the context 1364 for the Calibration References Data System. Particularly, residual fringes remain with the standard fringe removal, which could have an influence on weak spectral features (\citealt{Argyriou.etal.2020, Gasman.etal.2023}). Therefore, the ${\tt residual\_fringe}$ correction, disabled by default in the standard JWST pipeline, was activated to correct the fringe residuals (\citealt{Law.etal.2023}). In addition, the ${\tt imprint\_subtract}$ function was activated in JWST NIRSpec/IFU pipeline for leakage correction and the default ${\tt master\_bg}$ function was adopted in JWST MIRI/MRS pipeline for background subtraction. The JWST NIRSpec/IFU and MIRI/MRS pipelines ultimately produced one and twelve spectral data cubes for each target, respectively. The NIRSpec/IFU spectral data cube of each target has a FoV of $\sim 3\farcs4\times3\farcs4$, spanning the $2.87-5.27\,\mum$ wavelength range with the spectral resolution of $\sim 2000 - 4000$ (\citealt{Jakobsen.etal.2022, Shajib.etal.2025}). The twelve MIRI/MRS spectral data cubes of each target have FoVs ranging from $\sim 3\farcs6\times4\farcs5\,$ (ch1) to $\sim 7\farcs0\times8\farcs5\,$ (ch4) depending on the channel, together spanning the $4.90-27.90\,\mum$ wavelength range with the spectral resolution of $\sim 4000 - 2000$ (\citealt{Argyriou.etal.2023, Pontoppidan.etal.2024}).

\begin{figure*}[!ht]
\center{\includegraphics[width=0.95\linewidth]{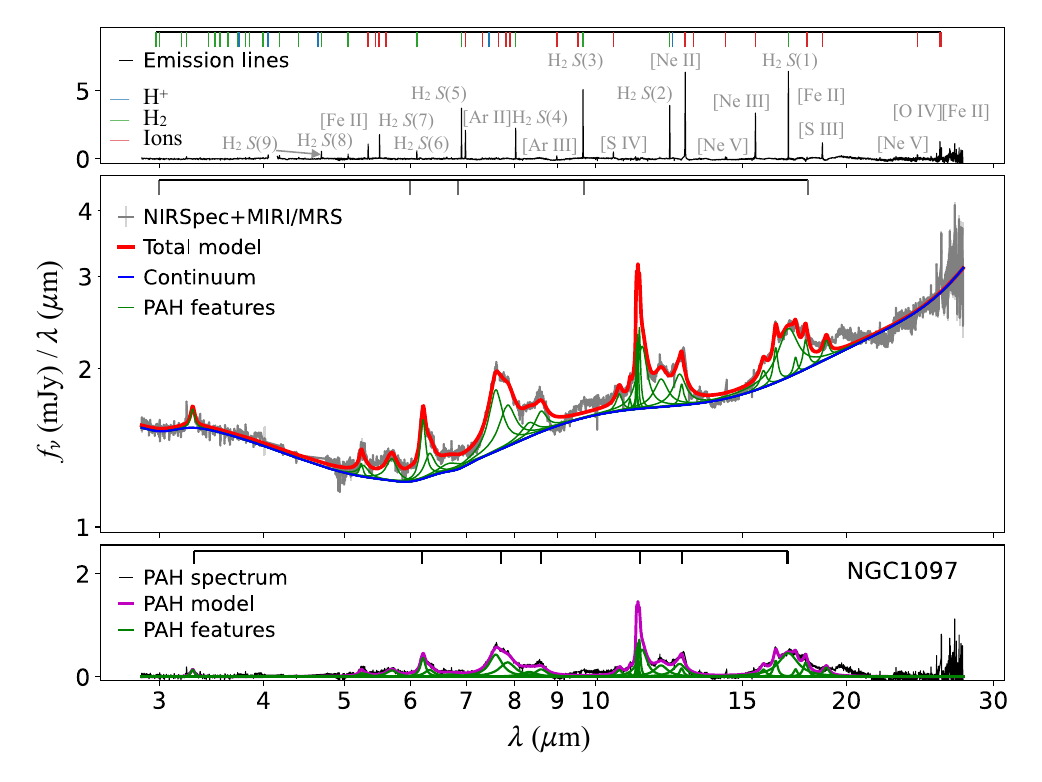}}
\caption{Top panel: The residual emission line spectrum obtained by subtracting the best-fit model (red curve in the middle panel) from the nuclear $\sim 3-28\,\mum$ spectrum. The positions of key emission lines (not necessarily all detected) are indicated by short-colored lines at the top. Middle panel: Illustrations of the multi-component fitting for PAH feature decomposition (see details in Section~\ref{sec2.3}), with the positions of the two silicate absorption features and the adopted ice absorption features for each target indicated by short-gray lines at the top. The gray curve is the nuclear NIRSpec/IFU + MIRI/MRS spectra, after masking the emission lines, as well as the CO$_2$ and CO features marked in panel b. The red curve shows the best-fit model, t he blue curve shows the sum of all continuum components (i.e., stellar and dust continuum), and these green curves are Drude profiles representing individual PAH features. Bottom panel: The residual PAH spectrum (black curve) obtained by subtracting all continuum components from the masked nuclear NIRSpec/IFU + MIRI/MRS spectrum. The positions of prominent PAH features around 3.3, 6.2, 7.7, 8.6, 11.3, 12.7, and 17.0 $\mum$ are indicated by short-black lines at the top. Same as in the middle panel, these green curves are Drude profiles representing individual PAH features, and the magenta curve shows the sum of all green curves, i.e., the modeled PAH spectrum. All x-axes and the y-axis of the middle panel are in logarithmic scale, while the other y-axes are in linear scale, and all spectra are in the rest frame. See Figures~\ref{DeSpec_II}, \ref{DeSpec_III}, and \ref{DeSpec_IV} in Appendix~\ref{secA0} for the same plots for NGC~1266, NGC~3190, and NGC~4736.}\label{DeSpec_I}
\end{figure*}

To extract spectra from the nuclear regions of the targets, we first characterized the unresolved AGN emission and estimated its wavelength-dependent contribution to the total flux within the central aperture. To this end, we fitted each slice in the data cubes with a model consisting of two two-dimensional Gaussian functions, one circular for the unresolved AGN component and the other allowed to be elliptical for the host galaxy, plus a uniform background. In this way, we obtained wavelength-dependent full widths at half maximum (FWHMs) for the AGN component that are consistent with those measured from JWST/MIRI MRS observations of point-like stars by \citet{Zhang&Ho2023}. More importantly, the unresolved AGN emission in all NIRSpec/IFU and MIRI/MRS bands were found to account for over 95\% of the total flux within radii equal to their FWHMs at all wavelengths for these targets. We then extracted the nuclear spectra using wavelength-dependent apertures from the full set of spectral data cubes for each target. The aperture center in each sub-band is the median position of the fitted AGN component within that sub-band, and the aperture radius is given by $r = 0.033\times\lambda + 0.106$ (e.g., $r = 0\farcs436$ at $\lambda = 10\,\mum$, corresponding to $r \lesssim 60$ pc at the distances of the four targets). The adopted aperture radius approximately equals to the FWHM value of the fitted point-spread-function-like (PSF-like) AGN component at each wavelength.

Subsequently, we applied the same wavelength-dependent aperture correction strategy to the extracted MIRI/MRS spectra as adopted by \cite{Gonzalez-Martin.etal.2025}, with the pre-computed aperture correction factors ($\sim 1.37 - 1.47$) based on in-flight observational MIRI/MRS PSFs.\footnote{Detailed information is included in the CRDS calibration file \href{https://jwst-crds.stsci.edu/browse/jwst_miri_apcorr_0008.asdf}{jwst\_miri\_apcorr\_0008.asdf}, and the wavelength-dependent aperture radius $r = 0.033\times\lambda + 0.106$ is also derived from this file.} Since no such calibration file exists for NIRSpec/IFU observations, we derived the aperture correction factors ($\sim 1.30 - 1.32$) from the encircled energy curves of theoretical NIRSpec/IFU PSFs computed by ${\tt Python}$ package ${\tt STPSF}$, and applied them for the extracted NIRSpec/IFU spectra. Finally, all aperture-corrected sub-band nuclear spectra were stitched together to produce a complete infrared spectrum spanning the $\sim 3-28\,\mum$ wavelength range for each target. Therein, systematic discontinuities, plausibly arising from data reduction uncertainties including aperture correction, between NIRSpec/IFU and MIRI/MRS sub-bands ($\lesssim 10\%$) and within some NIRSpec/IFU sub-bands ($\lesssim 5\%$) were corrected using constant scaling factors determined from the median fluxes in the overlapping or adjacent wavelength ranges.

\subsection{Spectral Decomposition and Measurement}\label{sec2.3}

\begin{figure*}[!ht]
\center{\includegraphics[width=1\linewidth]{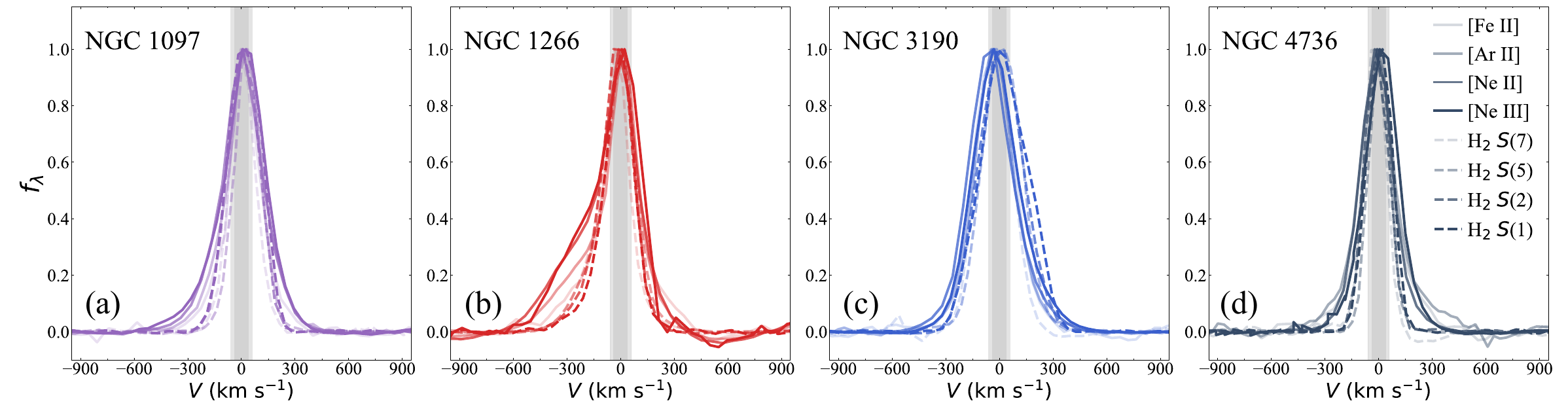}}
\caption{Normalized nuclear emission line profiles of [Fe~{\footnotesize II}], [Ar~{\footnotesize II}], [Ne~{\footnotesize II}], and [Ne~{\footnotesize III}], shown in order of increasing ionization potential from lighter to darker colors, along with H$_2\,S(7)$, H$_2\,S(5)$, H$_2\,S(2)$, and H$_2\,S(1)$, whose wavelengths are close to those of the ionized emission lines with the same colors, respectively. The gray shaded regions in each panel indicate the range of instrumental broadening (FWHM) at the wavelengths of these lines. All line profiles are shown in the rest frame.}\label{linepro}
\end{figure*}

As illustrated in Figure~\ref{DeSpec_I}, we first decomposed the PAH features from the extracted spectra through multi-component fitting, after masking all emission lines as well as the two adjacent CO$_2$ and CO features (of strong detection only in NGC~1266). The multi-component model consists of a series of Drude profiles for individual PAH features, multiple modified blackbodies for dust components, and a black body of 5000 $\rm K$ mimic the stellar continuum. Specifically, the Drude profiles with fixed widths are primarily adopted from \cite{Draine&Li2007} with some updates according to \cite{Lai.etal.2020}, \cite{Donnan.etal.2023}, and \cite{Draine.etal.2025}. The fixed temperatures of the modified blackbodies range from 35 to 1200 $\rm K$, in accordance with previous studies (i.e., \citealt{Smith.etal.2007, Donnan.etal.2023, Zhang.etal.2024b}). Moreover, all the model components are subject to foreground extinction, adopting the infrared extinction curve of \cite{Smith.etal.2007} (primarily for the silicate absorption features at 9.7 and 18 $\mum$), plus additional ice absorption features as reviewed by \cite{Boogert.etal.2015} (see Table~\ref{tabtaus} in Appendix~\ref{secB} for details of the adopted ice absorption features for each target). The decomposition was performed using the Bayesian Markov Chain Monte Carlo sampler $\tt emcee$ in the $\tt Python$ environment, with the median and standard deviation of the posterior distribution of each parameter adopted as the final estimate and its corresponding uncertainty (see Table~\ref{tabPAHs} in the Appendix~\ref{secB} for the PAH flux measurements). 

We then fitted all ionized emission lines, as well as the H$_2$ pure rotational (H$_2\,S(J)$, e.g., H$_2\,S(3)$) and ro-vibrational (H$_2$\,$\nu_{\rm up}$-$\nu_{\rm low}$\,$O(J)$, e.g., H$_2$\,1-0\,$O(5)$)\footnote{Only a few of the $O$-branch ro-vibrational transitions fall within the wavelength range covered by the nuclear spectra analyzed in this study.} transitions, individually with single- and double-Gaussian profiles plus a local linear continuum, using the Levenberg-Marquardt least-squares minimization algorithm. From visual inspection of the fits, we found that a double-Gaussian profile was required for all ionized emission lines, whereas a single-Gaussian profile was sufficient for the H$_2$ transitions. To obtain more robust statistics, each emission line spectrum was perturbed with random noise at the level of its uncertainty, and the fitting was repeated 100 times. The median and standard deviation of the 100 fits were adopted as the final flux estimate and its corresponding uncertainty for each emission line, respectively. Additionally, the fluxes of all emission lines were corrected for dust extinction using the extinction values derived from the multi-component full-spectrum fitting described above. See Tables~\ref{tablines} and \ref{tabH2s} in Appendix~\ref{secB} for the flux measurements of prominent ionized emission lines, as well as the H$_2$ pure rotational and ro-vibrational transitions detected in the nuclear spectra.

\section{Results}\label{sec3}

In this section, we characterize the emission from ionized gas, PAHs, and H$_2$ in the nuclear regions of the four low-luminosity AGN. By integrating insights from model results (\citealt{Guillard.etal.2012, Kristensen.etal.2023, Rigopoulou.etal.2024, Zhang.etal.2025}) and comparing with emission line and PAH measurements for nuclear regions and circumnuclear apertures of higher-luminosity Seyferts (log~$L^{\rm Xray}_{\rm (2-10)keV}$ = 40.9 -- 43.1; \citealt{Zhang.etal.2024a, Zhang.etal.2024b}), we find that shocks (i.e., the kinetic mode feedback) play a central role in shaping the nuclear environments of these low-luminosity AGN.

\subsection{Emission Line Profiles and Kinematics}\label{sec3.0}

As shown in Figures~\ref{DeSpec_I}, \ref{DeSpec_II}, \ref{DeSpec_III}, and \ref{DeSpec_IV}, the nuclear spectra of the four targets exhibit variations in continuum shape, extinction strength, PAH feature intensity (especially at shorter wavelengths) and in their emission line spectra. For instance, NGC~1266 exhibits the steepest dust-continuum slope (Figure~\ref{DeSpec_II}), more like the starburst-dominated Spitzer/IRS spectra, whereas the other three targets show v-shape dust continuums consistent with the LINER-dominated Spitzer/IRS spectra presented by \citeauthor{Gonzalez-Martin.etal.2015} (\citeyear{Gonzalez-Martin.etal.2015}; Figure~5 therein). NGC~1266 also shows the strongest extinction, relatively stronger short-wavelength PAH emission, and weaker high-ionization emission lines, implying the presence of unique evolutionary processes (see also e.g., \citealt{Davis.etal.2012, Pellegrini.etal.2013, Alatalo.etal.2014, Otter.etal.2024}). More detailed discussions of these variations are presented in the following subsections, while this subsection focuses on the kinematics of the ionized and molecular gas components revealed in the nuclear emission line spectra. Although a spatially resolved analysis, to be presented in subsequent papers, will better constrain the kinematic properties of these gas components, the kinematics of the emission lines in the nuclear spectra already provide important insights into the underlying feedback mechanisms.

Figure~\ref{linepro} shows the normalized profiles of four pairs of H$_2$ and ionized emission lines in the nuclear spectra, with each pair of emission lines having similar wavelengths. The key result in Figure~\ref{linepro} is that all ionized emission lines are generally broader than the paired H$_2$ lines, primarily due to the bilateral high-velocity components (especially those on the blue-shifted side) in ionized emission lines. Although this trend is less pronounced in NGC~3190 (see Figure~\ref{linepro}c), the ionized emission  lines in this galaxy are systematically blue-shifted relative to the H$_2$ emission lines. Overall, the widths of the ionized emission lines increase with ionization potential, a trend also observed in the nuclear regions of higher-luminosity Seyferts (e.g., \citealt{Armus.etal.2023, HermosaMunoz.etal.2024, Zhang.etal.2024a}). The markedly different kinematics of the ionized gas, particularly the highly ionized component, relative to H$_2$ imply the presence of ionized outflows launched from these nuclear regions, which we will examine in more detail in subsequent papers. The preliminary results in Figure~\ref{linepro} suggest that NGC~1266 exhibits the strongest ionized outflows among the four targets, whereas the weak outflow signature in NGC~3190 may be partly attributable to its almost edge-on orientation. More importantly, these results indicate that the ionized gas in the nuclear regions is, to some extent, decoupled from the molecular gas and appears to be more sensitive to AGN-driven outflows, as also discussed in our recent works (e.g., \citealt{Pereira-Santaella.etal.2022, Davies.etal.2024, AlonsoHerrero.etal.2025, Riffel.etal.2025}). This decoupling is important for unifying the different physical conditions around the AGN inferred from the ionized emission lines and H$_2$ transitions, as detailed in the following.

\begin{figure*}[!ht]
\center{\includegraphics[width=1\linewidth]{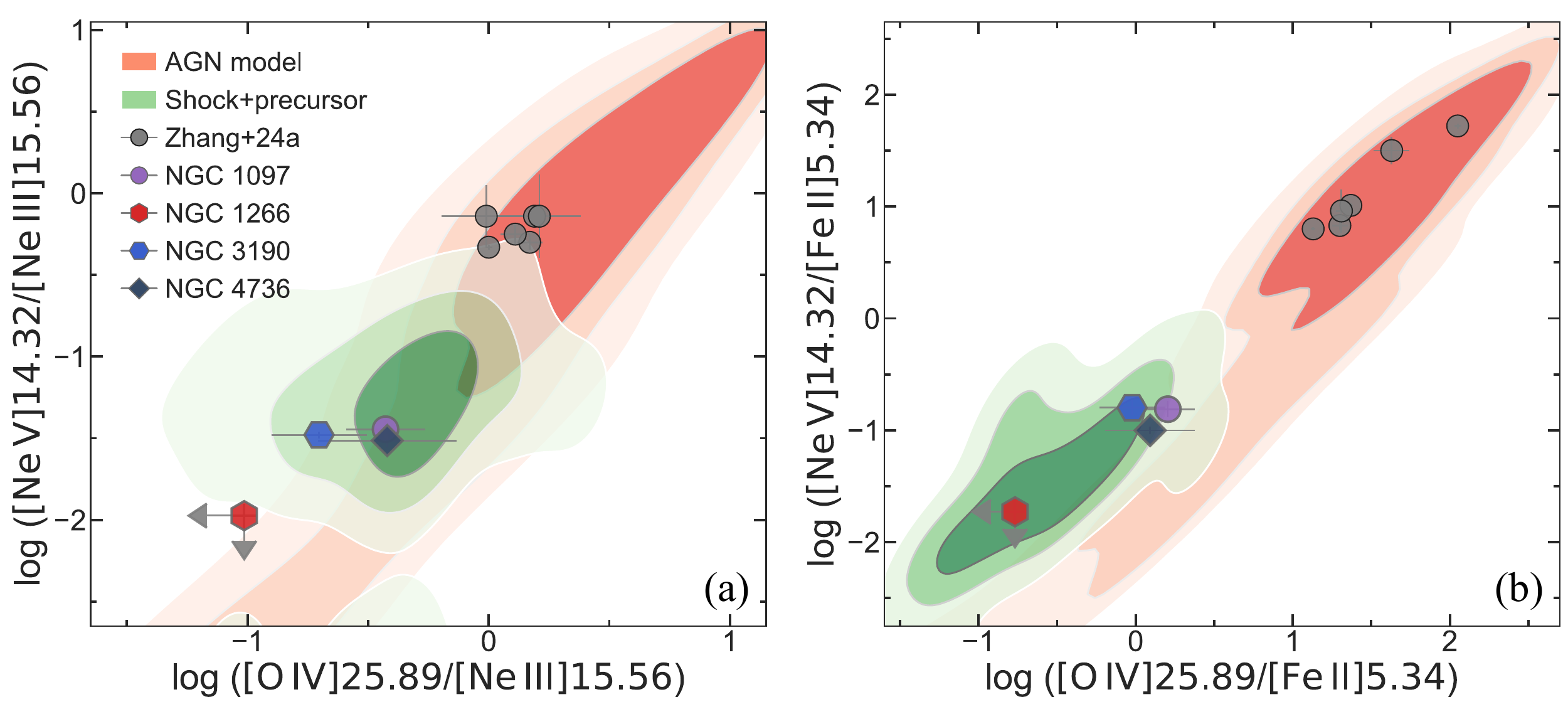}}
\caption{Diagnostic diagrams of ionized emission line ratios: (a) [Ne~{\footnotesize V}]/[Ne~{\footnotesize III}] versus [O~{\footnotesize IV}]/[Ne~{\footnotesize III}], and (b) [Ne~{\footnotesize V}]/[Fe~{\footnotesize II}] versus [O~{\footnotesize IV}]/[Fe~{\footnotesize II}]. The reddish and greenish contours show the distributions of model results computed by \cite{Zhang.etal.2025} for AGN and fast radiative shocks (including the shock precursor), respectively, with each contour enclosing 30\%, 60\%, and 90\% of the model results from inside to outside. The colored and gray data points represent measurements for the nuclear regions of the four low-luminosity AGN studied here and the six higher-luminosity Seyferts studied by \cite{Zhang.etal.2024a}, respectively. The line ratios of NGC~1266 are marked as upper limits because [Ne~{\footnotesize V}] and  [O~{\footnotesize IV}] flux measurements of this galaxy are below three times of the standard deviation noise of local continuum.}\label{Ion_diag}
\end{figure*}

\subsection{Ionized Emission Lines and Diagnostics}\label{sec3.1}

Ratios of infrared emission lines with different ionization potentials (IPs) and critical densities have been long proposed as powerful diagnostics of galaxy properties (e.g., \citealt{Spinoglio&Malkan1992, Genzel.etal.1998, Thornley.etal.2000, Dale.etal.2006, Armus.etal.2007, Pereira-Santaella.etal.2010, Pereira-Santaella.etal.2017, Feltre.etal.2023, Zhang.etal.2025}). Figure~\ref{Ion_diag} presents two diagnostic diagrams showing high-to-low ionization potential line ratios, along with the distributions of model results for AGN and fast radiative shocks. The AGN and fast radiative shock (including the shock precursor) models are from \cite{Zhang.etal.2025} computed using ${\tt MAPPINGS~V}$ (\citealt{Sutherland&Dopita2017}). Specifically, the AGN models shown in Figure~\ref{Ion_diag} adopt input parameters including the ionization parameter log~$U$, the AGN spectrum peak energy log~$(E_{\rm peak}/{\rm keV})$, the metallicity, and the gas pressure log~$(P/k)$, ranging from $-4.3$ to $-1.3$, $-2.1$ to $-0.9$, 0.2 to 2 $Z_{\odot}$, and 5.5 to 8.5, respectively. The shock models, in comparison, adopt input parameters including the shock velocity $v_s$, the pre-shock density $n_{\rm H}$, the magnetic to ram pressure ratio $\eta_{\rm M}$, and the metallicity, ranging from 100 to 500 $\rm km~s^{-1}$, 1 to $10^{4}$ $\rm cm^{-3}$, 0.0 to 0.1, and 0.2 to 2 $Z_{\odot}$, respectively (see  \citealt{Zhang.etal.2025} for more details).

Figure~\ref{Ion_diag} shows that the ionized gas in the four low-luminosity AGN is less highly excited, exhibiting weaker emission lines of high ionization potential compared to the nuclear $r = 0\farcs75$ regions of six higher-luminosity Seyferts (see also \citealt{Goold.etal.2024, AlonsoHerrero.etal.2025, HermosaMunoz.etal.2025a}). This trend is the most evident for NGC~1266, with the fluxes of [Ne~{\small V}]14.32$\mum$ (IP: 97.1~eV) and  [O~{\small IV}]25.89$\mum$ (IP: 54.9~eV) lines in the nuclear region of this galaxy below three times of the standard deviation noise of local continuum. The extremely weak high ionization potential lines in this post-starburst galaxy NGC~1266 is at least partially due to the highly obscured condition in the nuclear region of this galaxy, as indicated by the strong silicate and ice absorption features (see Figure~\ref{DeSpec_I}b). The highly obscured condition can be attributed to a gravitational encounter driving molecular gas into the nucleus (\citealt{Alatalo.etal.2014} and references therein). Upon close inspection of their nuclear spectra around specific wavelengths, we find that additional high ionization potential lines such as [Mg~{\small IV}]4.49$\mum$ (IP: 80.1~eV), [Mg~{\small V}]5.61$\mum$ (IP: 109.2~eV), and [Ne~{\small VI}]7.65$\mum$ (IP: 126.2~eV) are also detectable in the nuclear regions of NGC~1097 and NGC~3190, but not in NGC~1266 and NGC~4736, with the latter galaxy having log~$L^{\rm Xray}_{\rm (2-10)keV} < 40$. Interestingly, [Ar~{\small V}] lines at 7.90 and 13.10 $\mum$ (IP: 59.6~eV) are absent in the nuclear spectra of the four low-luminosity AGN, whereas [Ar~{\small VI}] line at 4.53 $\mum$ (IP: 74.8 eV) is tentatively detected in the nuclear spectra of NGC~1097 and NGC~3190. A comprehensive analysis of these ionized emission lines is deferred to a dedicated paper.

\begin{figure*}[!ht]
\center{\includegraphics[width=1\linewidth]{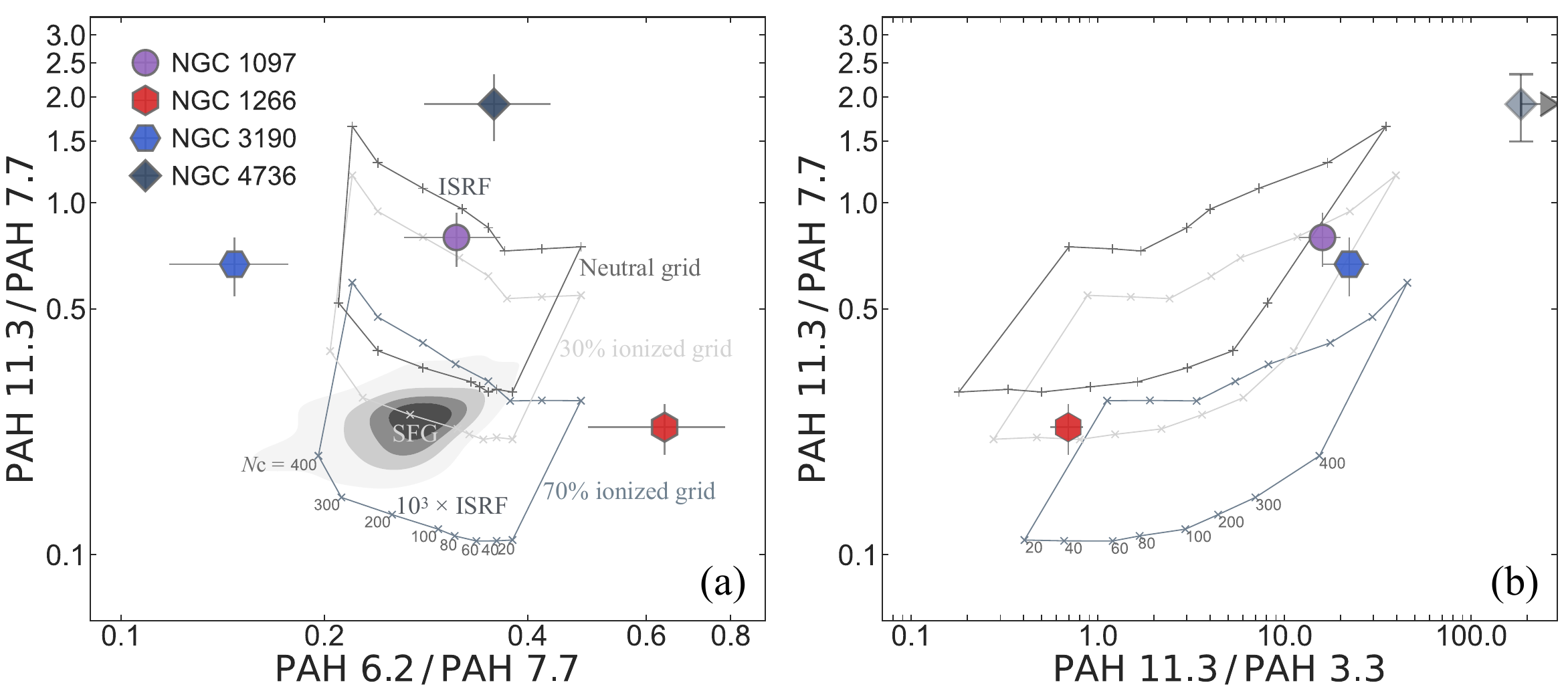}}
\caption{Diagnostic diagrams of PAH band ratios: (a) 11.3\,$\mum$/7.7\,$\mum$ versus 6.2\,$\mum$/7.7\,$\mum$ and (b) 11.3\,$\mum$/7.7\,$\mum$ versus 11.3\,$\mum$/3.3\,$\mum$, with the colored data points representing measurements in the nuclear regions of the four low-luminosity AGN. The gray contours represent the PAH band ratio distribution of spatially resolved regions in SINGS SFGs (\citealt{Zhang.etal.2022}). The gray grids represent model results of PAH band ratios by \cite{Rigopoulou.etal.2024} for neutral (top grid), 30\% ionized (middle grid), and 70\% ionized (bottom grid) PAHs of different sizes (with carbon number $N_{\rm C} = 20 - 400$ as marked in order), in the interstellar radiation field (ISRF; the top boundary of each grid) and the $10^3\times$ ISRF (the bottom boundary of each grid). Note that the 3.3 $\mum$ PAH feature in the nuclear region of NGC~4736 is not detected and the corresponding data point in panel (b) is shown as a faint upper limit with an arbitrary value for illustration purposes.}\label{PAH_diag}
\end{figure*}

In addition to the generally low-excitation conditions in these low-luminosity AGN, another key result shown by Figure~\ref{Ion_diag} is that the ionized gas in these sources appears to be primarily excited by fast ($\sim \rm 100s\,km\,s^{-1}$) radiative shocks, whereas the higher-luminosity Seyferts are dominated by AGN photoionization. This trend is especially evident in Figure~\ref{Ion_diag}(b), which involves [Fe~{\small II}]5.34$\mum$ line (IP: 7.9~eV) that is sensitive to shocks (e.g., \citealt{Forbes&Ward1993, Mouri.etal.2000, Koo.etal.2016}). We note that not only [Fe~{\small II}]5.34$\mum$ line, but also [Fe~{\small II}]17.93$\mum$ and [Fe~{\small II}]25.99$\mum$ lines are of detection in the four low-luminosity AGN, albeit the latter two lines in NGC~1266 are below three times of the standard deviation noise of local continuum. Nevertheless, the ionized gas in the nuclear region of NGC~1266 is plausibly dominated by shock excitation, as further discussed below and also revealed by optical and IR spectroscopy (\citealt{Davis.etal.2012, Pellegrini.etal.2013, Otter.etal.2024}). However, as will be discussed in Section~\ref{sec4}, even in the presence of shocks, additional physical conditions, such as a highly obscured environment or a strong AGN radiation, lead to variations in how AGN feedback affects the surrounding material in different galaxies.

\subsection{PAH Emission Features and Diagnostics}\label{sec3.2}

Theoretical calculations show that the size distribution and ionization state of PAH molecules, along with the input radiation field, play a key role in determining the relative strengths of individual PAH features (\citealt{Draine&Li2001, Draine&Li2007, Li&Draine2001, Maragkoudakis.etal.2020, Draine.etal.2021, Rigopoulou.etal.2021, Rigopoulou.etal.2024}). Specifically, smaller PAHs, upon absorbing a UV photon, reach much higher excitation levels and consequently radiate at shorter (more energetic) wavelengths because of their low heat capacity. Moreover, when excited, neutral PAHs exhibit stronger C--H bond vibrations responsible for 3.3 and 11.3 $\mum$ PAH features, whereas cationic PAHs show enhanced C--C bond vibrations that efficiently emit $6-9\,\mum$ PAH features. Furthermore, a moderately stronger radiation field results in a higher fraction of ionized PAHs, while an FUV-rich (i.e., harder) radiation field enhances the short-wavelength PAH features.

\begin{figure*}[!ht]
\center{\includegraphics[width=0.6\linewidth]{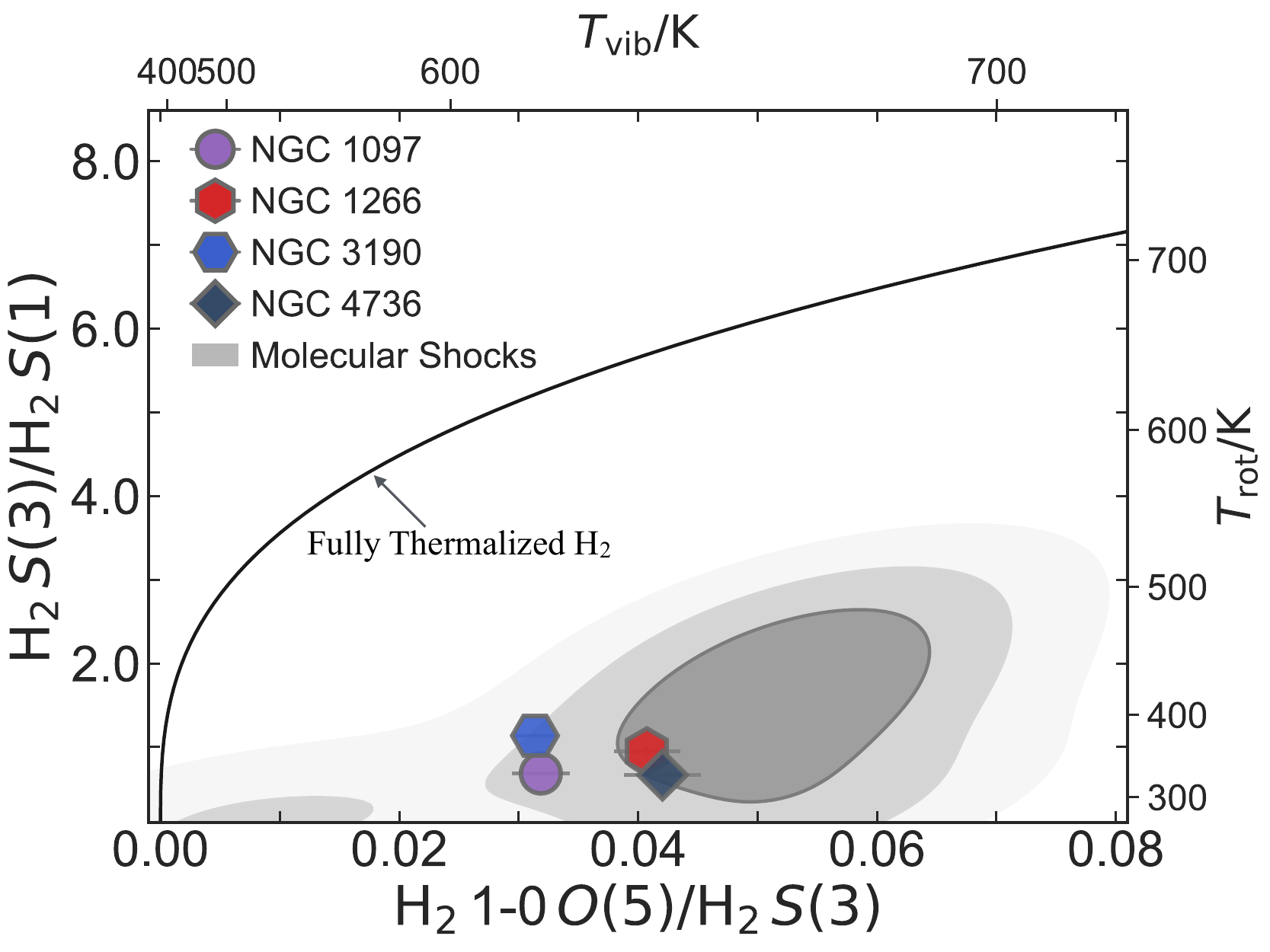}}
\caption{Diagnostic diagrams of H$_2$ transitions: H$_2\,S(3)$/H$_2\,S(1)$ versus H$_2$\,1-0\,$O(5)$/H$_2\,S(3)$, sensitive to H$_2$ rotational and vibrational temperatures, respectively. The colored data points represent measurements in the nuclear regions of the four low-luminosity AGN, while the gray contours show the distribution of the best-matched C-type molecular shock model results taken from \cite{Kristensen.etal.2023}. Additionally, the solid black curve represents the predicted trend of fully thermalized warm H$_2$ gas at different temperatures, with the corresponding $T_{\rm rot}$ and $T_{\rm vib}$ labeled on the right and upper axes, respectively.}\label{H2_diag}
\end{figure*}

Figure~\ref{PAH_diag} shows that the PAH characteristics in the nuclear regions of the four low-luminosity AGN are distinct from those of the spatially resolved regions in SFGs (\citealt{Zhang.etal.2022}). This finding clearly demonstrates the influence of AGN feedback, even under the low-excitation conditions prevailing in these low-luminosity AGN (see also \citealt{Zhang.etal.2022}). According to the model results of \cite{ Rigopoulou.etal.2024}, the nuclear regions of NGC~1097, NGC~3190, and NGC~4736 exhibit higher fractions of neutral PAHs with large sizes ($N_{\rm C} \gtrsim 200$), whereas the nuclear region of NGC~1266 shows a comparable fraction of ionized PAHs to those of SFGs, but with smaller sizes ($N_{\rm C} < 20$). The PAH characteristics in the nuclear region of NGC~1266 are broadly consistent with those of the highly obscured nuclei in NGC~3256, II~Zw~96, and VV~114 presented by \citeauthor{ Rigopoulou.etal.2024} (\citeyear{ Rigopoulou.etal.2024}; and references therein). The higher fractions of large, neutral PAHs in the nuclear regions of the other three targets are likely the result of the preferential destruction of the smaller, predominantly ionized PAHs, which are more susceptible to harsh environments surrounding AGN (e.g., \citealt{Allain.etal.1996b, Micelotta.etal.2010a, Micelotta.etal.2010b, Holm.etal.2011, Garcia-Bernete.etal.2022a, Garcia-Bernete.etal.2022b, Garcia-Bernete.etal.2024, Zhang.etal.2022, Zhang.etal.2024b, Zhang&Ho2023}). The primary cause of the preferential destruction appears to be the shocks discussed above and in the following subsection. Further discussion on the indications of the PAH characteristics in these three targets, also those in NGC~1266, are presented in Section~\ref{sec4}.

\subsection{Warm Molecular Gas Excitation}\label{sec3.3}

Ratios of pure rotational and ro-vibrational transitions from warm ($\sim 200 - 5000$ K) H$_2$ gas are also powerful diagnostics of the excitation mechanisms around AGN, offering key insights into the effects of AGN feedback (e.g., \citealt{Mouri1994, Roussel.etal.2007, Mazzalay.etal.2013, Kristensen.etal.2023, Lopez.etal.2025}). Figure~\ref{H2_diag} shows that warm H$_2$ gas in the nuclear regions of the four low-luminosity AGN is not fully thermalized, as indicated by the diagnostic ratios H$_2\,S(3)$/H$_2\,S(1)$ and H$_2$\,1-0\,$O(5)$/H$_2\,S(3)$ lying below the predicted curve for fully thermalized warm H$_2$ gas. Specifically, the rotational and vibrational temperatures of fully thermalized warm H$_2$ gas are given by $T_{\rm H_2} = -\frac{E_{i} - E_{j}}{k\,ln((\frac{F_{i}\lambda_{i}}{A_{i}g_{i}})/(\frac{F_{j}\lambda_{j}}{A_{j}g_{j}}))}$ (see Equations~\ref{equ1} and \ref{equ2}), using the flux ratios of H$_2\,S(3)$/H$_2\,S(1)$ and H$_2$\,1-0\,$O(5)$/H$_2\,S(3)$, and adopting the energy levels and other constants recently calculated by \cite{Roueff.etal.2019}. Furthermore, as shown by the gray contours in Figure~\ref{H2_diag}, the slow ($v_{\rm s} \leq \rm 10\,km\,s^{-1}$) C-type molecular shocks are plausibly the additional excitation source responsible for the deviation. Specifically, the best-matched molecular shock models are those of \cite{Kristensen.etal.2023}, characterized by shock velocity $v_{\rm s} \leq \rm 10\,km\,s^{-1}$, H atom density $n_{\rm H} \leq 10^{3}\,{\rm cm^{-3}}$, transverse magnetic field scaling factor $b = 1.0$, UV field strength $G_{0} = 0$, PAH abundance $X({\rm PAH}) = 10^{-6}$, and H$_2$ cosmic-ray excitation rate $\zeta_{\rm H_{2}} = 10^{-15}\,{\rm s^{-1}}$. The trend observed in Figure \ref{H2_diag} remains when the axes are replaced by the H$_2\,S(5)$/H$_2\,S(1)$ and H$_2$\,1-0\,$O(7)$/H$_2\,S(5)$ ratios, even though these ratios probe higher temperatures and require an external UV field for excitation (i.e., the best-matched H$_2$ shock models have $G_0 > 0$).

\noindent
\begin{align}\label{equ1}
\begin{aligned}
T_{\rm rot} \simeq \frac{-1489}{{\rm ln}\,(0.01748\times\frac{F_{{\rm H}_{2}\,S(3)}}{F_{{\rm H}_{2}\,S(1)}})}
\end{aligned}
\end{align}

\noindent
\begin{align}\label{equ2}
\begin{aligned}
T_{\rm vib} \simeq \frac{-4448}{{\rm ln}\,(0.02485\times\frac{F_{{\rm H}_{2}1-0\,O(5)}}{F_{{\rm H}_{2}\,S(3)}})}
\end{aligned}
\end{align}

\begin{figure*}[!ht]
\center{\includegraphics[width=1\linewidth]{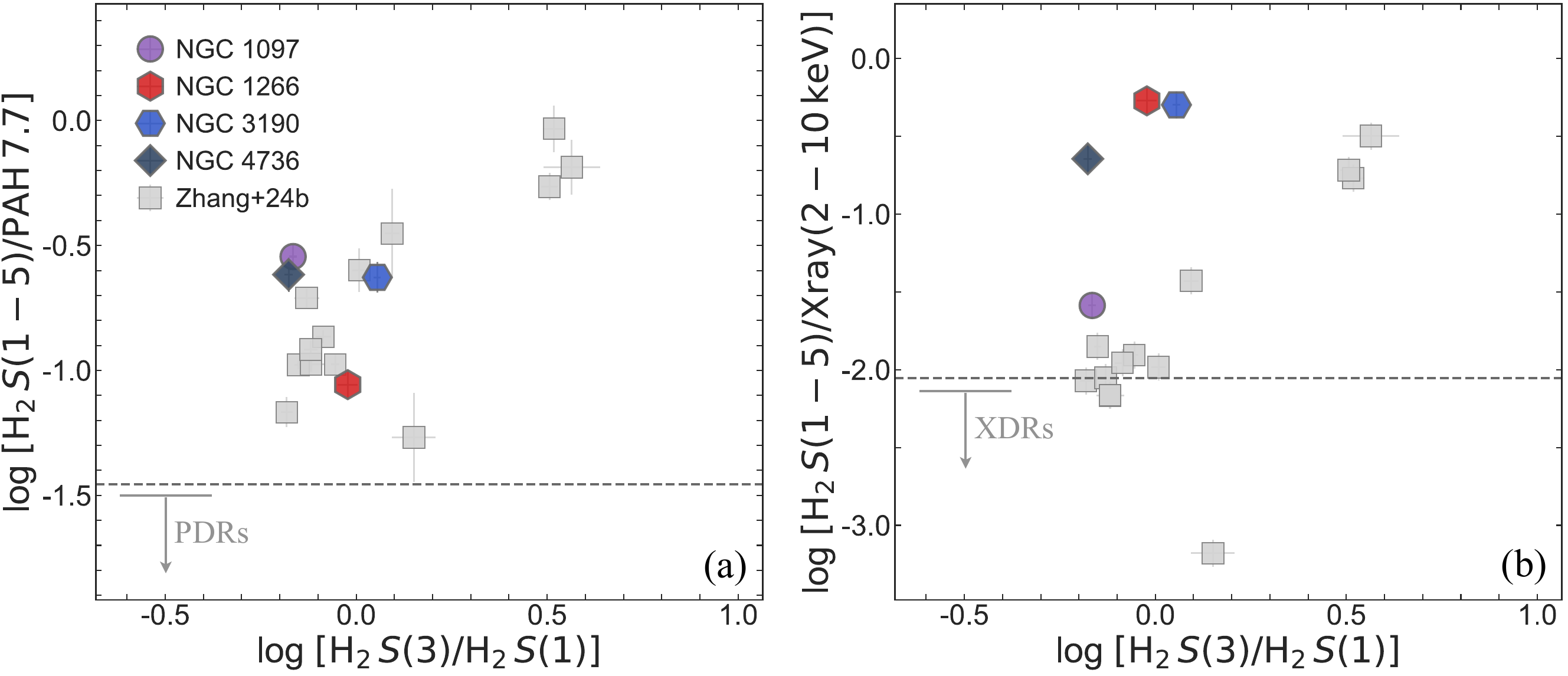}}
\caption{Diagnostic diagrams of (a) ${\rm H_{2}}\,S$(1--5)/PAH versus ${\rm H_{2}}\,S(3)/{\rm H_{2}}\,S(1)$ and (b) ${\rm H_{2}}\,S$(1--5)/X-ray versus ${\rm H_{2}}\,S(3)/{\rm H_{2}}\,S(1)$. The colored and gray data points represent measurements for the nuclear regions of the four low-luminosity AGN studied here and in the multiple apertures within the circumnuclear regions of three higher-luminosity Seyferts analyzed by \cite{Zhang.etal.2024b}, respectively. The horizontal dashed lines in panels (a) and (b) indicate the upper limits of ${\rm H_{2}}\,S$(1--5)/PAH\,7.7 and ${\rm H_{2}}\,S$(1--5)/X-ray(2--10 keV) that can be produced in PDRs and XDRs, respectively, as described in the main text (see Section~\ref{sec3.3}).}\label{HPX_diag}
\end{figure*}

Non-thermal UV heating in photodissociation regions and X-ray heating from AGN are also able to drive warm H$_2$ away from full thermalization; however, this is not the case here. Figure~\ref{HPX_diag} presents diagnostic diagrams showing the ratios of the summed H$_2$ $S$(1)--$S$(5) emission to the PAH 7.7 $\mum$ feature, ${\rm H_{2}}\,S$(1--5)/PAH7.7, and to the integrated 2--10 keV X-ray emission, ${\rm H_{2}}\,S$(1--5)/X-ray(2--10 keV), plotted against the H$_2$\,$S$(3)/H$_2$\,$S$(1) ratio. The measurements are shown for the nuclear regions of the four low-luminosity AGN and for thirteen $3\arcsec\times3\arcsec$ apertures sampling the circumnuclear regions of three higher-luminosity Seyfert galaxies. The corresponding upper limits that can be produced in photodissociation regions (PDRs) and X-ray--dominated regions (XDRs) are indicated by the horizontal dashed lines. Specifically, the upper limits are converted from the corresponding ${\rm H_{2}}\,S$(0--3)/PAH7.7 and ${\rm H_{2}}\,S$(0--3)/X-ray(2--10 keV) values (0.04 and 0.01) adopted by \cite{Guillard.etal.2012}, by multiplying a factor of 0.88. This conversion factor represents the average ${\rm H_{2}}\,S$(1--5)/${\rm H_{2}}\,S$(0--3) ratio predicted by the excitation models of H$_2$ gas at temperatures between 100 and 2000~K, calculated by \cite{Pereira-Santaella.etal.2014} using ${\tt RADEX}$. The ${\rm H_{2}}\,S$(1--5)/PAH7.7 and ${\rm H_{2}}\,S$(1--5)/X-ray(2--10 keV) ratios in the nuclear regions of the four low-luminosity AGN are all above the upper limits that can be produced in PDRs or XDRs, indicating that non-thermal UV and X-ray heating cannot account for the observed H$_2$ emission strength and thus further highlighting the impact of the slow molecular shocks. Additionally, as shown by the region with the lowest ${\rm H_{2}}\,S$(1--5)/X-ray(2--10 keV) values in Figure~\ref{HPX_diag}(b), and as will be further discussed in Section~\ref{sec4}, X-ray heating appears to play an important role in molecular gas excitation within the nuclear apertures of the higher-luminosity Seyferts that share PAH characteristics similar to those of quasars.

Figure~\ref{H2_diag} also shows that the warm H$_2$ gas, mostly traced by low-J pure-rotational transitions, in the nuclear regions of the four low-luminosity AGN does not reach particularly high temperatures. This result is consistent with a higher fraction of the H$_2$ gas distributed toward the lower-temperature end in these nuclear regions, as indicated by their higher power-law temperature distribution index $\beta$ ($\sim 5.0 - 5.5$; see Section~\ref{secA}), particularly when compared to the nuclear apertures of higher-luminosity Seyferts ($\beta \approx 4.0 - 5.0$; \citealt{Davies.etal.2024, Zhang.etal.2024b, Delaney.etal.2025}). This result further implies that non-thermal heating contributes to the warm H$_2$ excitation in the nuclear regions of the four low-luminosity AGN. As discussed above, slow molecular shocks are the most likely source of this non-thermal heating, and neither UV heating in PDRs nor X-ray heating from XDRs can account for the observed H$_2$ emission strength. Cosmic rays could be another heating source for the warm H$_2$ (\citealt{Ferland.etal.2008, Ogle.etal.2010}). However, even only for the warm H$_2$ content in these nuclear regions ($\sim 10^{4.2} - 10^{6.4}\,{\rm M_{\odot}}$; see Appendix~\ref{secA}), reproducing the observed H$_2$ fluxes through cosmic-ray heating alone would require cosmic-ray excitation rates of order $\sim10^{-11}\,{\rm s^{-1}}$, assuming an energy deposition of 20 eV per excitation event and $\sim$ 5\% of this energy ultimately going to H$_2$ rotational emission (i.e., $L_{\rm H_{2}}$ is of $N_{\rm H_{2}}\times\rm20\,eV\times0.05\times10^{-11}\,s^{-1}$). The implied cosmic-ray excitation rates are much higher than those expected for low-luminosity AGN ($\sim10^{-12}\,{\rm s^{-1}}$; \citealt{Koutsoumpou.etal.2025a,Koutsoumpou.etal.2025b}), indicating that cosmic-ray heating can only play a secondary role in powering the H$_2$ rotational emission in these nuclear regions.

In these low-luminosity AGN, molecular shocks, as the most likely heating source for the warm H$_2$, are in principle driven by radio jets launched from the nuclear regions (\citealt{Mukherjee.etal.2018, Meenakshi.etal.2022}). Quantitatively, a coupling of only $\sim 0.1 - 1$\% of the jet power ($L_{\rm jet}$) is sufficient to reproduce the observed H$_2$ fluxes in most of these nuclear regions. The exception is NGC~1266, which requires a coupling of $\sim 4$\%. To estimate $L_{\rm jet}$, we adopt the empirical correlation between $L_{\rm jet}$ and $L^{\rm radio}_{\rm 1.4GHz}$ described by \citeauthor{Heckman&Best2014} (\citeyear{Heckman&Best2014}; Equation~1 therein with $f_W = 15$). The estimated couplings remain largely unchanged when using the updated $L_{\rm jet}$ calibration provided by \citeauthor{Roy.etal.2025} (\citeyear{Roy.etal.2025}; Equation 5 therein). Interestingly, the coupling efficiencies estimated here for the warm H$_2$ emission are comparable to those inferred for ionized gas outflows in high-redshift radio galaxies ($< 1$\%; \citealt{Roy.etal.2025}) and in NGC~1266 ($\sim 4$\%; \citealt{Nyland.etal.2013}). This result further highlights the important and multifaceted roles played by the radio (kinetic) mode AGN feedback in galaxy evolution.

\section{Discussion: Insights into AGN Feedback Mechanisms}\label{sec4}

\begin{figure*}[!ht]
\center{\includegraphics[width=0.8\linewidth]{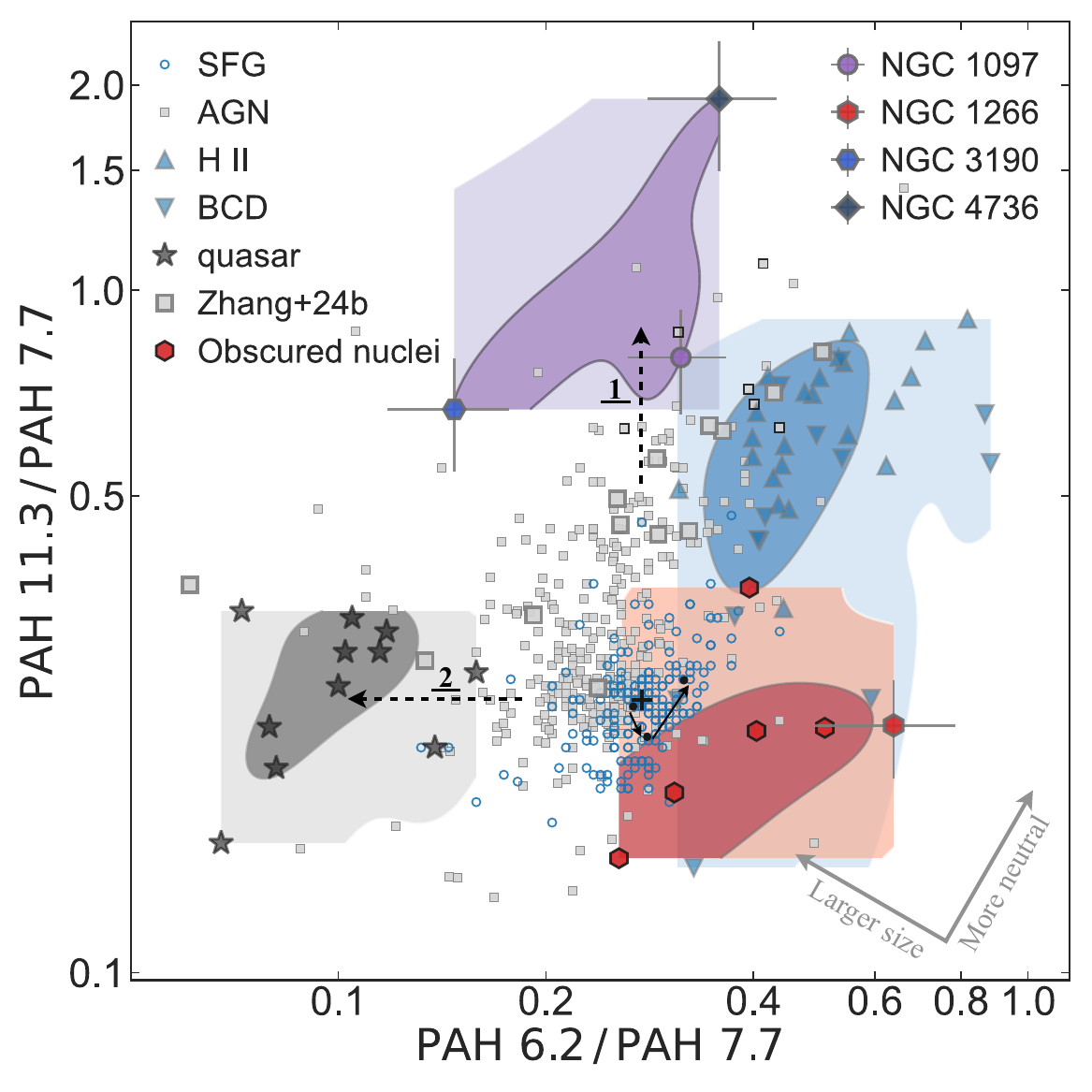}}
\caption{The same PAH band ratio diagram as Figure~\ref{PAH_diag}(a), but including additional data points compiled from the literature. Blue circles: the distribution of PAH band ratios of spatially resolved regions in SINGS SFGs (\citealt{Zhang.etal.2022}), with the black plus representing the median value of all SFG regions and the black dots (from left to right) indicating the median values of SFG regions with log\,[Ne~{\scriptsize III}]/[Ne~{\scriptsize II}] $< -0.7$, $-0.7 \leq$ log\,[Ne~{\scriptsize III}]/[Ne~{\scriptsize II}] $\leq -0.5$, and log\,[Ne~{\scriptsize III}]/[Ne~{\scriptsize II}] $> -0.5$, respectively. Small gray squares: PAH band ratios of spatially resolved circumnuclear regions in SINGS AGN, and central regions of a larger set of AGN. The small gray squares further outlined in black are Seyfert central regions with the most pronounced H$_2$ emission in \cite{Diamond-Stanic&Rieke2010}. Gray stars: PAH band ratios of low-redshift quasars detected in all the three PAH features. Blue triangles: PAH band ratios of low-metallicity BCDs and local giant H~{\footnotesize II} regions. Large gray squares: PAH band ratios of the multiple apertures sampling the circumnuclear regions of three Seyferts as also shown in Figure~\ref{HPX_diag} (\citealt{Zhang.etal.2024b}). Red hexagons: PAH band ratios of highly obscured nuclei studied by \cite{ Rigopoulou.etal.2024}, together with the nuclei of NGC~1266 analyzed here. Other markers: the nuclei of the other three low-luminosity AGN analyzed in this work. Dashed black arrows: the two potential branches of AGN feedback ({\underline 1}: kinetic mode; {\underline 2}: radiative mode). Gray arrows in the lower-right: the directions in theory of increasing overall PAH size and neutral fraction as the PAH band ratios change. Colored contours: the density contours of the overall distributions of different galactic environments. Note that the PAH band ratios from \cite{Rigopoulou.etal.2024}, \cite{Zhang.etal.2024b}, and this work were obtained from higher–spatial-resolution JWST/MRS spectroscopy, whereas those from the other studies shown here are based on Spitzer/IRS spectroscopy.}\label{PAH_diag_plus}
\end{figure*}

PAHs in space are not a single, homogeneous population but instead span a broad molecular size distribution, with carbon number of $\sim 10 - 1000$ (e.g., \citealt{Li&Draine2001, Weingartner&Draine2001, Draine.etal.2021}). Broadly speaking, the size distribution and ionization stage of PAH population in space are governed by their formation and subsequent processing, and variations in the observed PAH band ratios therefore reflect the underlying physical processes acting on PAH population. PAHs can form via both ``bottom-up'' pathways, in which PAHs grow from smaller molecules (e.g., \citealt{Frenklach&Feigelson1989, Cherchneff.etal.1992}; and see the review by \citealt{Reizer.etal.2022}), and ``top-down'' pathways, in which PAHs are produced through the fragmentation of larger grains (e.g., \citealt{Jones.etal.1996, Scott.etal.1997, Raj.etal.2014, Hirashita&Aoyama2019, Hirashita&Murga2020}). Subsequent PAH processing includes photo-processing (e.g., \citealt{Jochims.etal.1994, Allain.etal.1996a, Holm.etal.2011}) and shock-processing (e.g., \citealt{Jones.etal.1996, Micelotta.etal.2010a, Micelotta.etal.2010b}). Consequently, by combining emission models for individual PAH sub-populations, analyses of variations in observed PAH band ratios can provide insight into the physical processes governing PAH formation and processing in different environments. To this end, Figure~\ref{PAH_diag_plus} revisits the widely used PAH band-ratio diagram, incorporating additional PAH measurements from the literature to sample a broad range of PAH band ratios and, in turn, the underlying physical conditions. Note that the additional PAH measurements come from diverse galactic environments observed with comparable arcsec-scale aperture sizes but were derived using the same profile-fitting methodology, ensuring a consistent and systematic comparison.

The literature PAH measurements compiled here are not intended to be exhaustive across all galactic systems, but rather to sample typical star-forming and AGN central regions, along with a few distinctive environments that exhibit unusual PAH characteristics, in order to illustrate the sensitivity of PAH features to differing physical conditions, as reflected in their distinct overall PAH band ratios. Accordingly, the added galactic systems span spatially resolved regions from the central areas of SINGS SFGs and AGN (including both LINERs and Seyferts; \citealt{Zhang.etal.2022}), central-region observations from a larger set of AGN (most are Seyferts; \citealt{Smith.etal.2007, ODowd.etal.2009, Diamond-Stanic&Rieke2010, Gallimore.etal.2010, Sales.etal.2010, Garcia-Bernete.etal.2022b}), low-redshift quasars (\citealt{Xie&Ho2022}), low-metallicity blue compact dwarf galaxies (BCDs; \citealt{Hunt.etal.2010, Lebouteiller.etal.2011}), local giant H~{\footnotesize II} regions in NGC~3603 (in the Milky Way), 30~Doradus (in the Large Magellanic Cloud), and N~66 (in the Small Magellanic Cloud) (\citealt{Lebouteiller.etal.2011}), as well as multiple apertures within the circumnuclear regions of three Seyferts with JWST spectroscopy (\citealt{Zhang.etal.2024b}), and highly obscured nuclei with JWST spectroscopy (\citealt{ Rigopoulou.etal.2024}). As noted above, the distinct distributions of these galactic systems on the PAH diagram plausibly arise from differing feedback effects that regulate the interstellar medium (ISM) properties. Despite the distinct PAH characteristics, their connection through the SFG and AGN measurements hints at a more continuous and interconnected evolutionary pathway of the ISM and, consequently, of galaxies. The following discussion provides a more comprehensive view of the feedback effects, especially those in AGN, by integrating the information conveyed by the PAH characteristics and other diagnostics discussed above for these galactic systems.

SFGs exhibit similar PAH characteristics, as evidenced by the concentrated distribution of the spatially resolved regions in SFGs in Figure~\ref{PAH_diag_plus}, and thus serve as a useful baseline for highlighting the effects of different feedback mechanisms. Within the clustered distribution, the PAH characteristics of SFG regions are primarily regulated by the strength and hardness of their radiation fields ( \citealt{Zhang.etal.2022}). Specifically, \cite{Zhang.etal.2022} found that the average PAH 6.2~$\mum$/7.7~$\mum$ ratio of SFG regions increases with increasing [Ne~{\small III}]/[Ne~{\small II}], an indicator of radiation field hardness (e.g., \citealt{Thornley.etal.2000, ForsterSchreiber.etal.2001}), while the average PAH 11.3~$\mum$/7.7~$\mum$ ratio first decreases with [Ne~{\small III}]/[Ne~{\small II}] and then the trend reverses. The first decrease of PAH 11.3~$\mum$/7.7~$\mum$ ratio is consistent with a higher fraction of ionized PAHs and enhanced short-wavelength PAH features in a moderately strong, hard radiation field. Additionally, \cite{Zhang.etal.2022} proposed that an excessively strong and hard radiation field causes selective photo-erosion and subsequent photo-destruction of PAHs, primarily the ionized ones, shifting the PAH size distribution toward smaller and more neutral molecules (the ``top-down'' scenario). These photo-processing effects explain the final increase of both PAH 6.2~$\mum$/7.7~$\mum$ and PAH 11.3~$\mum$/7.7~$\mum$ ratios along the direction toward the distribution of the low-metallicity BCDs and local giant H~{\footnotesize II} regions in Figure~\ref{PAH_diag_plus}.

Following the trend seen in SFGs, the photo-processing effects in the even stronger and harder radiation fields of low-metallicity BCDs and local giant H~{\footnotesize II} regions at least partially explain their on average large values of both PAH 6.2~$\mum$/7.7~$\mum$ and PAH 11.3~$\mum$/7.7~$\mum$ ratios. Furthermore, as detailed by \cite{Whitcomb.etal.2024,Whitcomb.etal.2025}, the inhibited growth of PAHs, which successfully reproduces the relatively elevated (but absolutely declined) fraction of the smallest PAHs among the overall PAH population in the low-metallicity environments, can further explain the distribution of these BCDs and H~{\footnotesize II} regions (the ``bottom up'' scenario). The selective erosion and the inhibited growth of PAHs may also apply to the highly obscured nuclei in Figure~\ref{PAH_diag_plus}, given their elevated fractions of smaller PAHs. However, in these highly obscured nuclei, scenarios involving the selective erosion of larger PAHs, or, alternatively, the enhanced production of smaller PAHs through fragmentation of large grains, are particularly favored, as evidenced by their not only relatively but also absolutely stronger PAH 3.3\,$\mum$ feature predominantly tracing the smallest PAH population. Specifically, NGC~1266 nucleus exhibits strong absorption features from organic residues around 7.42 and 7.674 $\mum$. The cold, dense environments  in these highly obscured nuclei provide the ideal physical conditions for dust growth and molecular synthesis (see review \citealt{Herbst&vanDishoeck2009}). The highly obscured, ice-rich environments may also protect ionized PAHs from photo-destruction (e.g., \citealt{Alonso-Herrero.etal.2014, Alonso-Herrero.etal.2020, Garcia-Bernete.etal.2022b, Garcia-Bernete.etal.2025}), explaining the large fraction of very small yet ionized PAHs observed therein. Also notably, the host galaxies of these highly obscured nuclei exhibit peculiar morphologies, indicative of gravitational interactions capable of channeling molecular gas toward their nuclei. We thus suspect that the molecular shocks responsible for the enhanced H$_2$ emission and the associated selective erosion/fragmentation of PAHs may be the relics of past gravitational encounters.

In contrast to the concentrated distribution of SFG regions, AGN central regions occupy a much broader distribution of PAH band ratios, displaying characteristically larger 11.3~$\mum$/7.7~$\mum$ but smaller 6.2~$\mum$/7.7~$\mum$ toward {\it two} potential branches. The PAH characteristics observed in AGN can be attributed to the destruction of smaller, predominantly ionized PAHs by feedback effects, though the underlying mechanisms appear to differ between low- and high-luminosity AGN. As detailed in Section~\ref{sec3}, in low-luminosity AGN toward the {\it vertical} branch, the feedback mechanism responsible for the destruction is most likely shock-driven, associated with the kinetic mode feedback (see also \citealt{Zhang.etal.2022}). However, in low-redshift quasars toward the {\it horizontal} branch, photo-destruction by their intense radiation fields is favored given that these systems are dominated by radiative-mode feedback. In line with the latter point, the higher-luminosity Seyferts whose nuclear apertures show non-negligible X-ray heating of the molecular gas as shown by Figure \ref{HPX_diag}(b) also exhibit PAH characteristics similar to those of the quasars in Figure~\ref{PAH_diag_plus} (see more discussion in \citealt{Zhang.etal.2024b}). Moreover, X-ray photons penetrate much deeper into molecular gas than UV photons, allowing them to irradiate and excite large PAH molecules that would otherwise remain neutral. As a result, the surviving large PAHs in quasars are predominantly ionized, whereas those in most Seyferts remain mostly neutral. For other AGN central regions that lie between the SFG loci and those of low-luminosity AGN and quasars, a combination of shocks, intense AGN radiation, and star formation, may be required to explain their PAH characteristics. This interpretation is particularly relevant for AGN central regions observed with larger apertures, or for obscured AGN where host-galaxy inclination or a dense nuclear gas disk contributes to the observed properties. More importantly, although based on a modest sample of low-luminosity AGN, the preliminary analysis suggests that SFG regions, low-luminosity AGN, and quasars collectively form a useful set of anchor points for assessing the relative influence of different AGN feedback modes through a mixing-sequence PAH diagram. A more rigorous quantitative analysis will require spatially resolved JWST spectroscopy of an expanded AGN sample, particularly at the low-luminosity end.

Another noteworthy result in Figure~\ref{PAH_diag_plus} is that the nuclear apertures of higher-luminosity Seyferts with the most pronounced H$_2$ emission, an indicator of shocks, in Figure~\ref{HPX_diag}, as well as the central regions of Seyferts with the most pronounced H$_2$ emission in \cite{Diamond-Stanic&Rieke2010}, occupy a region between the low-luminosity AGN and low-metallicity environments. Most of these Seyferts also exhibit peculiar morphologies and could be analogs of the above highly obscured nuclei, but with higher AGN luminosities (though still below those dominated by X-ray photo-processing effects). If this supposition is correct, the distribution of these Seyferts with stronger signatures of molecular shocks and a relatively higher fraction of small PAHs could result from the competition between different feedback effects. These effects include the selective erosion/fragmentation of larger PAH grains, producing more smaller PAHs (as seen in the highly obscured nuclei), and the complete destruction of smaller PAHs, either by shocks (as in the nuclear regions of the low-luminosity AGN) or by the AGN radiation (as in quasars). Galactic systems that exhibit signatures of both AGN activity and gravitational interactions, such as some ultraluminous infrared galaxies (ULIRGs), will be ideal laboratories to test this idea in future studies.

\section{Summary and Conclusions}\label{sec5}

Leveraging high quality JWST NIRSpec/IFU and MIRI/MRS spectroscopy obtained as an extension of the Galaxy Activity, Torus, and Outflow Survey (GATOS), this letter presents the distinct emission properties of ionized gas, PAHs, and H$_2$ in the nuclear ($r < 150$ pc) regions of a sample of four low-luminosity AGN. We find that these low-luminosity AGN exhibit much weaker high-ionization potential lines (e.g., [Ne~{\small V}] and [O~{\small IV}]) compared to higher-luminosity Seyferts, and have emission line ratios indicative of fast radiative shocks (with $v_{\rm s}$ of $\sim \rm 100s\,km\,s^{-1}$) as the primary excitation source of ionized gas therein (Section~\ref{sec3.1}). Despite their low-excitation conditions, their PAH characteristics point to a critical role of AGN feedback in regulating the ISM properties of these low-luminosity AGN (Section~\ref{sec3.2}). Furthermore, the H$_2$ gas in these low-luminosity AGN are not fully thermalized, with slow, plausibly jet-driven molecular shocks (with $v_{\rm s} \leq \rm 10\,km\,s^{-1}$) likely responsible for its excitation. The role of molecular shocks is further supported by the fact that non-thermal UV, X-ray, and cosmic-ray heating are all insufficient to explain the observed H$_2$ emission strength, whereas radio jets with a coupling efficiency of $\sim 0.1 - 4$\% can reproduce the observed H$_2$ fluxes in these nuclear regions. (Section~\ref{sec3.3}).

Taken together with results from the literature, we specifically discussed the underlying mechanisms responsible for the widely distributed PAH band ratios observed across different galactic systems, particularly among AGN of varying luminosities (Section~\ref{sec4}). The main points of the discussion can be summarized as follows:

\begin{enumerate}

\item In SFGs, a moderately strong and hard radiation field leads to a higher fraction of ionized PAHs and enhanced emission features at short wavelengths, whereas an excessively strong and hard radiation field causes selective photo-erosion and subsequent photo-destruction of PAHs, shifting the PAH size distribution toward relatively smaller and more neutral molecules.

\item In AGN, feedback effects tend to destroy smaller, predominantly ionized PAHs, though the underlying mechanisms differ with AGN luminosity. In low-luminosity AGN, the destruction is likely driven by shocks associated with the prevalent kinetic-mode feedback therein, whereas in quasars that are dominated by radiative-mode feedback, photo-destruction by the intense AGN radiation is favored. For most AGN, the mixed effect of shocks and AGN radiation fields is required.

\item Certain galactic systems such as low-metallicity BCDs and local giant H~{\footnotesize II} regions, highly obscured nuclei, and some Seyferts exhibiting the most prominent H$_2$ emission, appear to contain an elevated fraction of smaller PAHs, either ionized or neutral depending on the physical conditions. The underlying processes responsible for the production of these smaller PAHs involve photo-processing, inhibited grain growth, enhanced PAH formation, or the combination thereof.

\end{enumerate}

Although based on a modest sample, our analysis of the infrared ($\sim 3-28\,\mum$) spectra indicates that shocks (i.e., the kinetic mode feedback) play a central role in shaping the nuclear environments of low-luminosity AGN. Moreover, SFGs, low-luminosity AGN, and quasars together provide a useful set of anchor points for assessing the relative influence of different AGN feedback modes through a mixing-sequence PAH diagram. A more robust, quantitative analysis will require further spatially resolved analysis and JWST spectroscopy of an expanded AGN sample in future cycles, particularly at the low-luminosity end, as well as distinctive systems such as Seyferts with strong H$_2$ emission and ULIRGs exhibiting signatures of both AGN activity and gravitational interactions.


\acknowledgements
We thank the anonymous referee for detailed comments and suggestions to improve the presentation of our results and corresponding discussions. This letter is part of a series from the \href{https://gatos.myportfolio.com}{Galaxy Activity, Torus, and Outflow Survey (GATOS) collaboration}. LZ and CP acknowledge grant support from the Space Telescope Science Institute (ID: JWST-GO-01670; JWST-GO-03535; JWST-GO-04225; JWST-GO-04972). EKSH and DD acknowledge support from the NASA Astrophysics Data Analysis Program (22-ADAP22-0173). MPS acknowledges support from grants RYC2021-033094-I, CNS2023-145506, and PID2023-146667NB-I00 funded by MCIN/AEI/10.13039/501100011033 and the European Union NextGenerationEU/PRTR. CR acknowledges support from SNSF Consolidator grant F01$-$13252, Fondecyt Regular grant 1230345, ANID BASAL project FB210003 and the China-Chile joint research fund. AAH and LHM acknowledge financial support by the grant PID2021-124665NB-I00 funded by MCIN/AEI/10.13039/501100011033 and ERDF A way of making Europe. IGB is supported by the Programa Atracci\'on de Talento Investigador ``C\'esar Nombela'' via grant 2023-T1/TEC-29030 funded by the Community of Madrid. EB acknowledges support from the Spanish grants PID2022-138621NB-I00 and PID2021-123417OB-I00, funded by MCIN/AEI/10.13039/501100011033/FEDER, EU. SFH acknowledges support through UK Research and Innovation (UKRI) under the UK government’s Horizon Europe Funding Guarantee (EP/Z533920/1, selected in the 2023 ERC Advanced Grant round) and an STFC Small Award (ST/Y001656/1).


\appendix

\section{PAH Decomposition}\label{secA0}
\begin{figure*}[!ht]
\figurenum{A1}
\center{\includegraphics[width=0.9\linewidth]{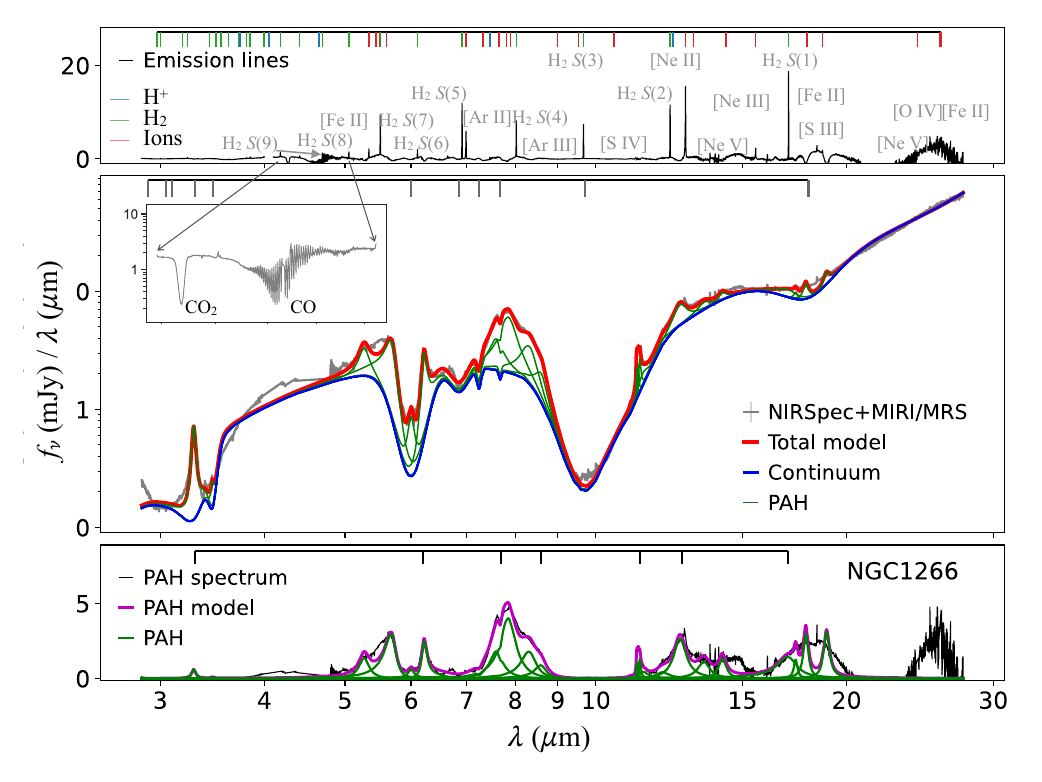}}
\caption{The same as Figure~\ref{DeSpec_I} but for NGC~1266.}\label{DeSpec_II}
\end{figure*}

\begin{figure*}[!ht]
\figurenum{A2}
\center{\includegraphics[width=0.9\linewidth]{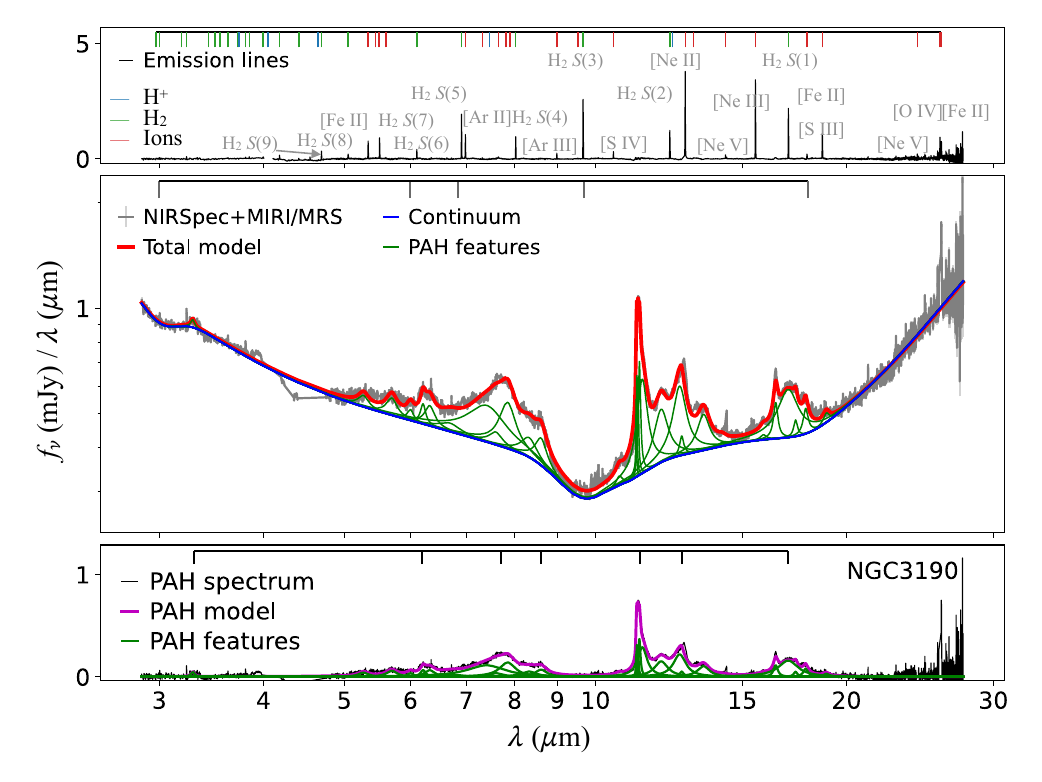}}
\caption{The same as Figure~\ref{DeSpec_I} but for NGC~3190.}\label{DeSpec_III}
\end{figure*}

\begin{figure*}[!ht]
\figurenum{A3}
\center{\includegraphics[width=0.9\linewidth]{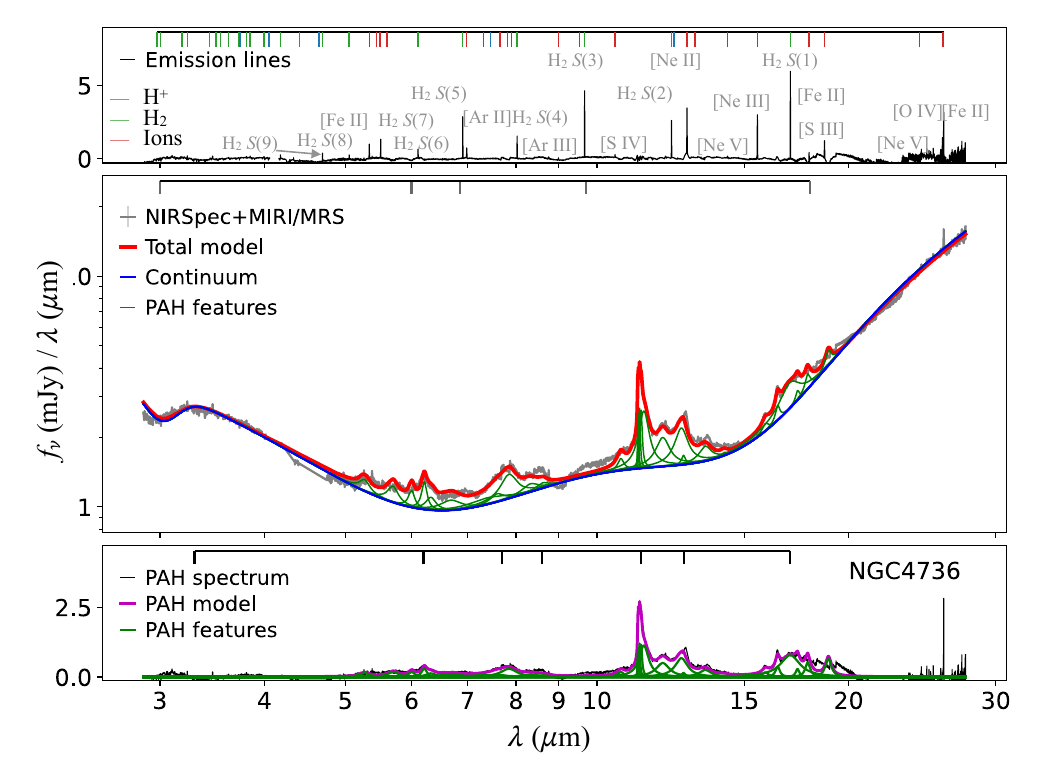}}
\caption{The same as Figure~\ref{DeSpec_I} but for NGC~4736.}\label{DeSpec_IV}
\end{figure*}

Figures~\ref{DeSpec_II}, \ref{DeSpec_III}, and \ref{DeSpec_IV} for PAH decompositions of NGC~1266, NGC~3190, and NGC~4736.

\newpage

\newpage

\newpage

\section{H$_2$ Population Diagrams}\label{secA}

The flux of an H$_2$ transition at energy level $J$, $F(J)$, is related to the H$_2$ column density in the upper level, $N(J_{\rm up})$, by $F(J) \propto A \times N(J_{\rm up})/\lambda$, where $A$ and $\lambda$ are the Einstein coefficient and wavelength of the corresponding H$_2$ transition, respectively.

\begin{figure*}[!ht]
\figurenum{A4}
\center{\includegraphics[width=0.9\linewidth]{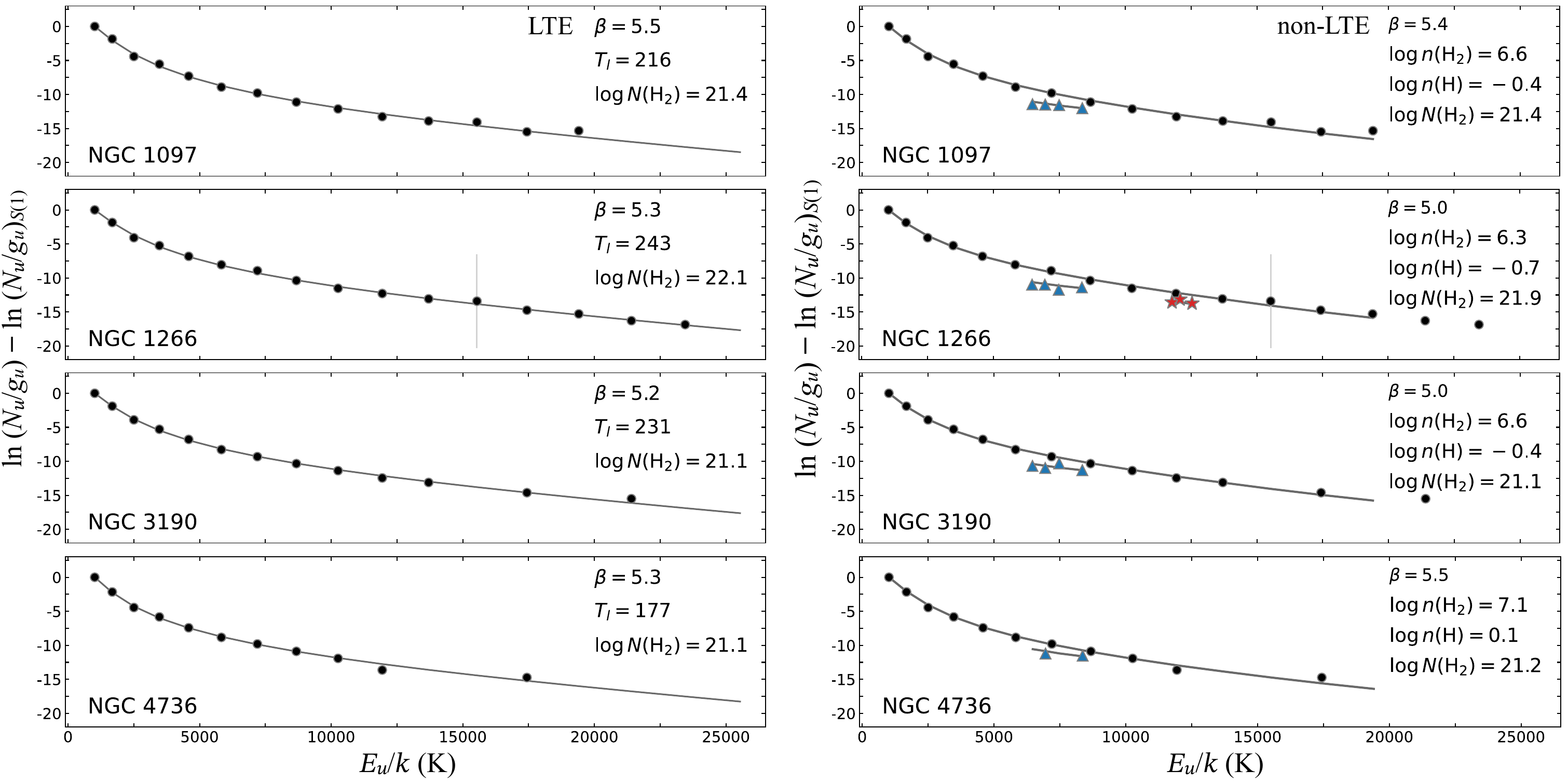}}
\caption{H$_2$ population diagrams of (left) the pure rotational transitions and (right) the pure rotational and ro-vibrational transitions. Black dots in each panel represent the relative populations of the H$_2$ pure rotational transitions, H$_2\,S(1)$ through H$_2\,S(16)$, arranged from left to right if detected. Blue triangles in right panels represent the relative populations of the H$_2$ ro-vibrational transitions, H$_2$\,1-0\,$O(4)$ through H$_2$\,1-0\,$O(7)$, arranged from left to right if detected. Red stars in the right panel of NGC~1266 represent the relative populations of the H$_2$ ro-vibrational transitions, H$_2$\,2-1\,$O(3)$,  H$_2$\,2-1\,$O(4)$, and H$_2$\,2-1\,$O(5)$, arranged from left to right. The black curves show the predicted population distributions based on the best-fit parameters labeled in the top right of each panel, assuming LTE for the H$_2$ pure rotational transitions in the left panels and non-LTE conditions for the H$_2$ pure rotational and ro-vibrational transitions in the right panels, as detailed in Section~\ref{secA}. Note that the H$_2$ column density ($N_{\rm H_2}$) labeled in each panel is derived assuming an equivalent aperture size of $1\arcsec \times 1\arcsec$ for each nuclear region.}\label{H2pop}
\end{figure*}

Assuming local thermal thermodynamic (LTE) and a realistic power-law distribution of the H$_2$ level populations as a function of temperature $T$, i.e., $dN = mT^{-\beta}dT$, with $m = {N_{\rm tot}(\beta-1)}/({T_{l}^{1-\beta} - T_{u}^{1-\beta}}$), the column density of each H$_2$ pure rotational transition can be expressed as $N(J) = m\int_{T_{l}}^{T_{u}}\frac{g(J)}{Z(T)} \, e^{-E(J)/kT}T^{-\beta}dT$ (\citealt{Togi&Smith2016}; see also \citealt{Hunt.etal.2025}). This relation is fitted to the observed relative populations of the H$_2$ pure rotational transitions to derive the free parameters (see the left panels of Figure~\ref{H2pop}). Following \cite{Togi&Smith2016}, we fit for the lower-end temperature ($T_{l}$) and the power-law index ($\beta$), while fixing the upper-end temperature at $T_{u} = 3500\,{\rm K}$. The adopted $T_{u}$ is higher than that used by \cite{Togi&Smith2016} ($T_{u} = 2000\,{\rm K}$), since the fitting here includes H$_2$ pure rotational transitions from H$_2\,S(1)$ through H$_2\,S(16)$, whereas \cite{Togi&Smith2016} considered transitions from H$_2\,S(0)$ through H$_2\,S(7)$. The best-fit values of $T_{l}$ and $\beta$, together with the corresponding total H$_2$ column density $N({\rm H_2})$ under LTE, are shown in the upper right corners of the left panels in Figure~\ref{H2pop}.

The LTE assumption does not hold here when the fitting includes both H$_2$ pure rotational and ro-vibrational transitions, although the power-law distribution of H$_2$ level populations as a function of temperature remains applicable. In this case, the column density of each H$_2$ pure rotational or ro-vibrationa transition can be expressed as $N(J) = m\int_{T_{l}}^{T_{u}}n(T, n({\rm H_2}), n({\rm H}))T^{-\beta}dT$ (\citealt{Pereira-Santaella.etal.2014}), where $m$ is defined as above, and $n(T, n({\rm H_2}), n({\rm H}))$ represents the fractional H$_2$ population at level $J$ at given temperature for an environment with H$_2$ and H number densities of $n({\rm H_2})$ and $n({\rm H})$, respectively. This relation is fitted to the observed relative populations of the H$_2$ pure rotational and ro-vibrational transitions using $n(T, n({\rm H_2}), n({\rm H}))$ numerically calculated by \cite{Pereira-Santaella.etal.2014} (available with $J$ up to 14) to derive the free parameters (see the right panels of Figure~\ref{H2pop}). Following \cite{Pereira-Santaella.etal.2014}, we fit for $\beta$, $n({\rm H_2})$ and $n({\rm H})$, while fixing the lower- and upper-end temperatures at $T_{l} = 200\,{\rm K}$ and $T_{u} = 3500\,{\rm K}$. The best-fit values of $\beta$, $n({\rm H_2})$ and $n({\rm H})$, together with the corresponding total H$_2$ column density ($N({\rm H_2})$) under non-LTE conditions, are shown in the upper right corners of the right panels in Figure~\ref{H2pop}.

\section{Absorption and Emission Measurements}\label{secB}

Tables for measurements of silicate and ice absorptions (Table~\ref{tabtaus}), PAH features (Table~\ref{tabPAHs}),  ionized emission lines (Table~\ref{tablines}), and H$_2$ transitions (Table~\ref{tabH2s}) involved in this work.

\startlongtable
\tablenum{A1}
\setlength{\tabcolsep}{6pt}
\begin{deluxetable*}{cccccc}
\tablecolumns{6}
\tablecaption{Measurements of Silicate and Ice Absorptions}
\tablehead{
\colhead{Component} & \colhead{FWHM} & \colhead{log $\tau^{\rm NGC~1097}$} & \colhead{log $\tau^{\rm NGC~1266}$} & \colhead{log $\tau^{\rm NGC~3190}$} & \colhead{log $\tau^{\rm NGC~4736}$}\\
\colhead{(-)} & \colhead{($\mum$)} & \colhead{(-)}  & \colhead{(-)}  & \colhead{(-)}  & \colhead{(-)} \\
\colhead{(1)} & \colhead{(2)} & \colhead{(3)} & \colhead{(4)} & \colhead{(5)} & \colhead{(6)} }
\startdata
Silicate$^{\dagger}$ & $\cdots$ & $-$4.98 &  \,\,\,\,0.45 & $-$0.58 & $-$5.01 \\
Ice$_{\rm 3.0\,\mum}^{\ddagger}$ & 0.353 & $-$1.43 & $\cdots$ & $-$0.88 & $-$0.48 \\
$_{-\rm 2.90\,\mum}$ & 0.101 & $\cdots$ & $-$4.97 & $\cdots$ & $\cdots$ \\
$_{-\rm 3.05\,\mum}$ & 0.294 & $\cdots$ & $-$0.44 & $\cdots$ & $\cdots$ \\
$_{-\rm 3.10\,\mum}$ & 0.212 & $\cdots$ & $-$1.31 & $\cdots$ & $\cdots$ \\
$_{-\rm 3.30\,\mum}$ & 0.294 & $\cdots$ & \,\,\,\,0.09 & $\cdots$ & $\cdots$ \\
Ice$_{\rm 3.47\,\mum}$ & 0.094 & $\cdots$ & $-$0.09 & $\cdots$ & $\cdots$ \\
Ice$_{\rm 6.0\,\mum}$ & 0.577 & $-$1.83 &  \,\,\,\,0.32 & $-$4.99 & $-$5.02 \\
Ice$_{\rm 6.85\,\mum}$ & 0.398 & $-$1.97 & $-$0.31 & $-$4.99 & $-$5.00 \\
Ice$_{\rm 7.24\,\mum}$ & 0.099 & $\cdots$ & $-$0.42 & $\cdots$ & $\cdots$ \\
Ice$_{\rm 7.674\,\mum}$ & 0.066 & $\cdots$ & $-$0.76 & $\cdots$ & $\cdots$ \\
\enddata
\tablecomments{Optical depths of silicate and ice absorptions. [$\dagger$]: The infrared extinction curve of \cite{Smith.etal.2007} is given by the optical depth $\tau(\lambda)$ normalized by the peak value at 9.7~$\mum$, $\tau(9.7\mum)$, as listed in the first row. [$\ddagger$]: The extinction of each ice absorption component is given by the optical depth $\tau(\lambda)$ modeled with a Gaussian profile using a fixed FWHM and a fitted amplitude as listed.}
\label{tabtaus}
\end{deluxetable*}

\startlongtable
\tablenum{A2}
\setlength{\tabcolsep}{10pt}
\begin{deluxetable*}{ccccc}
\tablecolumns{5}
\tablecaption{Measurements of PAH Features}
\tablehead{
\colhead{PAH features} & \colhead{log $f_{\rm PAH}^{\rm NGC~1097}$} & \colhead{log $f_{\rm PAH}^{\rm NGC~1266}$} & \colhead{log $f_{\rm PAH}^{\rm NGC~3190}$} & \colhead{log $f_{\rm PAH}^{\rm NGC~4736}$}\\
\colhead{(-)} & \colhead{[erg/s/cm$^2$]}  & \colhead{[erg/s/cm$^2$]}  & \colhead{[erg/s/cm$^2$]}  & \colhead{[erg/s/cm$^2$]} \\
\colhead{(1)} & \colhead{(2)} & \colhead{(3)} & \colhead{(4)} & \colhead{(5)} }
\startdata
PAH$_{\rm 3.3\,\mum}$ & $-$14.09$\pm$0.09 & $-$12.10$\pm$0.08 & $-$14.46$\pm$0.10 & $\cdots$ \\
PAH$_{\rm 6.2\,\mum}$ & $-$13.30$\pm$0.05 & $-$11.82$\pm$0.08 & $-$13.77$\pm$0.07 & $-$13.34$\pm$0.07 \\
PAH$_{\rm 7.7\,\mum}$ & $-$12.79$\pm$0.05 & $-$11.62$\pm$0.06 & $-$12.94$\pm$0.06 & $-$12.90$\pm$0.06 \\
PAH$_{\rm 11.3\,\mum}$ & $-$12.89$\pm$0.06 & $-$12.26$\pm$0.04 & $-$13.12$\pm$0.06 & $-$12.61$\pm$0.07 \\
\enddata
\tablecomments{Flux measurements of PAH features.}
\label{tabPAHs}
\end{deluxetable*}

\startlongtable
\tablenum{A3}
\setlength{\tabcolsep}{5pt}
\begin{deluxetable*}{ccccccc}
\tablecolumns{7}
\tablecaption{Measurements of Ionized Emission Lines}
\tablehead{
\colhead{Ionized lines} & \colhead{$\lambda$} & \colhead{IP} & \colhead{log $f_{\rm NGC~1097}$} & \colhead{log $f_{\rm NGC~1266}$} & \colhead{log $f_{\rm NGC~3190}$} & \colhead{log $f_{\rm NGC~4736}$}\\
\colhead{(-)} & \colhead{($\mum$)} & \colhead{(eV)} & \colhead{[erg/s/cm$^2$]}  & \colhead{[erg/s/cm$^2$]}  & \colhead{[erg/s/cm$^2$]}  & \colhead{[erg/s/cm$^2$]} \\
\colhead{(1)} & \colhead{(2)} & \colhead{(3)} & \colhead{(4)} & \colhead{(5)} & \colhead{(6)} & \colhead{(7)} }
\startdata
$\rm[Fe~II]$ & 5.34 & 7.9 & $-$14.60$\pm$0.06 & $-$14.06$\pm$0.02 & $-$14.64$\pm$0.00 & $-$14.63$\pm$0.03 \\
$\rm[Ar~II]$ & 6.985 & 15.8 & $-$14.24$\pm$0.03 & $-$13.53$\pm$0.01 & $-$14.50$\pm$0.02 & $-$14.74$\pm$0.01 \\
$\rm[Ar~III]$ & 8.991 & 27.6 & $-$14.99$\pm$0.02 & $-$15.08$\pm$0.07 & $-$15.14$\pm$0.15 & $-$15.45$\pm$0.10 \\
$\rm[S~IV]$ & 10.511 & 34.8 & $-$14.84$\pm$0.07 & $-$15.50$\pm$0.13 & $-$15.06$\pm$0.02 & $-$15.37$\pm$0.09 \\
$\rm[Ne~II]$ & 12.814 & 21.6 & $-$13.70$\pm$0.04 & $-$13.10$\pm$0.01 & $-$13.89$\pm$0.03 & $-$13.96$\pm$0.01 \\
$\rm[Ne~V]$ & 14.322 & 97.1 & $-$15.41$\pm$0.05 & $<-$15.79 & $-$15.42$\pm$0.01 & $-$15.63$\pm$0.02 \\
$\rm[Ne~III]$ & 15.555 & 41.0 & $-$13.97$\pm$0.03 & $-$13.82$\pm$0.03 & $-$13.94$\pm$0.02 & $-$14.11$\pm$0.04 \\
$\rm[S~III]$ & 18.71 & 23.3 & $-$14.37$\pm$0.11 & $-$14.12$\pm$0.21 & $-$14.36$\pm$0.10 & $-$14.45$\pm$0.08 \\
$\rm[O~IV]$ & 25.89 & 54.9 & $-$14.39$\pm$0.16 & $<-$14.83 & $-$14.65$\pm$0.20 & $-$14.54$\pm$0.28 \\
\enddata
\tablecomments{Flux measurements of ionized emission lines.}
\label{tablines}
\end{deluxetable*}

\startlongtable
\tablenum{A4}
\setlength{\tabcolsep}{6pt}
\begin{deluxetable*}{cccccc}
\tablecolumns{6}
\tablecaption{Measurements of H$_2$ Transitions}
\tablehead{
\colhead{H$2$ transitions} & \colhead{$\lambda$} & \colhead{log $f_{\rm NGC~1097}$} & \colhead{log $f_{\rm NGC~1266}$} & \colhead{log $f_{\rm NGC~3190}$} & \colhead{log $f_{\rm NGC~4736}$}\\
\colhead{(-)} & \colhead{($\mum$)} & \colhead{[erg/s/cm$^2$]}  & \colhead{[erg/s/cm$^2$]}  & \colhead{[erg/s/cm$^2$]}  & \colhead{[erg/s/cm$^2$]} \\
\colhead{(1)} & \colhead{(2)} & \colhead{(3)} & \colhead{(4)} & \colhead{(5)} & \colhead{(6)} }
\startdata
H$_2\,S(1)$ & 17.035 & $-$13.80$\pm$0.01 & $-$13.23$\pm$0.02 & $-$14.12$\pm$0.01 & $-$13.97$\pm$0.01 \\
H$_2\,S(2)$ & 12.279 & $-$14.06$\pm$0.01 & $-$13.50$\pm$0.02 & $-$14.41$\pm$0.01 & $-$14.37$\pm$0.01 \\
H$_2\,S(3)$ & 9.665 & $-$13.96$\pm$0.01 & $-$13.26$\pm$0.02 & $-$14.07$\pm$0.01 & $-$14.14$\pm$0.01 \\
H$_2\,S(4)$ & 8.025 & $-$14.35$\pm$0.01 & $-$13.65$\pm$0.02 & $-$14.56$\pm$0.01 & $-$14.64$\pm$0.01 \\
H$_2\,S(5)$ & 6.91 & $-$14.16$\pm$0.01 & $-$13.39$\pm$0.03 & $-$14.26$\pm$0.01 & $-$14.38$\pm$0.01 \\
H$_2\,S(6)$ & 6.109 & $-$14.94$\pm$0.01 & $-$14.00$\pm$0.03 & $-$14.99$\pm$0.01 & $-$15.07$\pm$0.01 \\
H$_2\,S(7)$ & 5.511 & $-$14.51$\pm$0.01 & $-$13.57$\pm$0.03 & $-$14.63$\pm$0.01 & $-$14.68$\pm$0.01 \\
H$_2\,S(8)$ & 5.053 & $-$15.27$\pm$0.02 & $-$14.38$\pm$0.02 & $-$15.26$\pm$0.01 & $-$15.34$\pm$0.02 \\
H$_2\,S(9)$ & 4.6947 & $-$14.98$\pm$0.02 & $-$14.15$\pm$0.05 & $-$14.98$\pm$0.01 & $-$15.06$\pm$0.01 \\
H$_2\,S(10)$ & 4.4096 & $-$15.73$\pm$0.04 & $-$14.74$\pm$0.05 & $-$15.70$\pm$0.04 & $-$16.06$\pm$0.11 \\
H$_2\,S(11)$ & 4.181 & $-$15.34$\pm$0.02 & $-$14.41$\pm$0.02 & $-$15.32$\pm$0.01 & $\cdots$ \\
H$_2\,S(12)$ & 3.9947 & $-$15.72$\pm$0.21 & $-$14.87$\pm$3.00 & $-$14.90$\pm$0.07 & $\cdots$ \\
H$_2\,S(13)$ & 3.8464 & $-$15.70$\pm$0.03 & $-$14.81$\pm$0.03 & $-$15.65$\pm$0.02 & $-$15.54$\pm$0.07 \\
H$_2\,S(14)$ & 3.724 & $-$15.98$\pm$0.05 & $-$15.40$\pm$0.04 & $\cdots$ & $\cdots$ \\
H$_2\,S(15)$ & 3.625 & $\cdots$ & $-$15.23$\pm$0.03 & $-$15.78$\pm$0.05 & $\cdots$ \\
H$_2\,S(16)$ & 3.547 & $\cdots$ & $-$15.86$\pm$0.05 & $\cdots$ & $\cdots$ \\
H$_2$\,1-0\,$O(4)$ & 3.0039 & $-$15.85$\pm$0.06 & $-$15.09$\pm$0.01 & $-$15.85$\pm$0.06 & $\cdots$ \\
H$_2$\,1-0\,$O(5)$ & 3.235 & $-$15.43$\pm$0.03 & $-$14.65$\pm$0.02 & $-$15.53$\pm$0.02 & $-$15.48$\pm$0.05 \\
H$_2$\,1-0\,$O(6)$ & 3.5007 & $-$16.01$\pm$0.07 & $-$15.51$\pm$0.04 & $-$15.78$\pm$0.05 & $\cdots$ \\
H$_2$\,1-0\,$O(7)$ & 3.8075 & $-$15.83$\pm$0.03 & $-$14.99$\pm$0.02 & $-$15.84$\pm$0.03 & $-$15.78$\pm$0.03 \\
H$_2$\,2-1\,$O(3)$ & 2.9741 & $\cdots$ & $-$15.61$\pm$0.03 & $\cdots$ & $\cdots$ \\
H$_2$\,2-1\,$O(4)$ & 3.1899 & $\cdots$ & $-$15.87$\pm$0.05 & $\cdots$ & $\cdots$ \\
H$_2$\,2-1\,$O(5)$ & 3.4379 & $\cdots$ & $-$15.68$\pm$0.03 & $\cdots$ & $\cdots$ \\
\enddata
\tablecomments{Flux measurements of H$_2$ transitions.}
\label{tabH2s}
\end{deluxetable*}

\,\,\,


\end{document}